\documentstyle[12pt,psfig,myplots,aaspp]{article}
\begin{document}

\textwidth 6.5in
\textheight 8.5in
\topmargin 0.0in
\oddsidemargin 0.0in

\title{Transient Lunar Phenomena: Regularity and Reality}

\author{Arlin P.S.~Crotts}
\affil{Department of Astronomy, Columbia University,
Columbia Astrophysics Laboratory,\\
550 West 120th Street, New York, NY 10027}

\begin{abstract}
Transient lunar phenomena (TLPs) have been reported for centuries, but their
nature is largely unsettled, and even their existence as a coherent phenomenon
is still controversial.
Nonetheless, a review of TLP data shows regularities in the observations; a key
question is whether this structure is imposed by human observer effects,
terrestrial atmospheric effects or processes tied to the lunar surface.

I interrogate an extensive catalog of TLPs to determine if human
factors play a determining role in setting the distribution of TLP reports.
We divide the sample according to variables which should produce varying
results if the determining factors involve humans e.g., historical epoch or
geographical location of the observer, and not reflecting phenomena tied to the
lunar surface.
Specifically, we bin the reports into selenographic areas (300 km on a side),
then construct a robust average count for such ``pixels'' in a way discarding
discrepant counts.
Regardless of how we split the sample, the results are very similar: roughly
50\% of the report count originate from the crater Aristarchus and vicinity,
$\sim$16\% from Plato, $\sim$6\% from recent, major impacts (Copernicus, Kepler
and Tycho - beyond Aristarchus), plus a few at Grimaldi.
Mare Crisium produces a robust signal for three of five averages of up to 7\%
of the reports (however, Crisium subtends more than one pixel).
The consistency in TLP report counts for specific features on this list indicate
that $\sim$80\% of the reports are consistent with being real (perhaps with the
exception of Crisium).

Some commonly reported sites disappear from the robust averages, including
Alphonsus, Ross D and Gassendi.
TLP reports supporting these sites originate almost entirely after year 1955,
when TLPs became more popular targets of observation and many more (and
inexperienced) observers searched for TLPs.

We review non-lunar hypotheses discussed to explain TLP, and find conflicts
with the data involving nearly all of them.
Furthermore, in a companion paper, we compare the spatial distribution of
robust TLP sites of transient outgassing (seen by instruments on Apollo and
{\it Lunar Prospector}).
To a high confidence against the random hypothesis, robust TLP sites and those
of lunar outgassing correlate strongly, further arguing for the reality of TLPs.
\end{abstract}

\medskip
\section{Introduction}

TLPs (Transient Lunar Phenomena, called LTPs by some authors) as we are
describing them, are seen at optical wavelengths, typically during visually
observations through a telescope (sometimes photographically, discussed below).
There is no commonly accepted physical explanation for TLPs, and some authors
even question if they are due to processes local to the Moon at all.
Cameron (1972) divides TLPs (from a catalog of 771 reported events) into four
categories:
``brightenings:'' white or color-neutral increases in surface brightness,
``reddish:'' red, orange or brown color changes with or without brightening,
``bluish:'' green, blue or violet color changes with or without brightening,
and ``gaseous:'' obscurations, misty or darkening changes in surface
appearance.
Nearly all TLPs are highly localized, usually to a radius much less than
100 km, often as unresolved points (corresponding to roughly 1 km or less).
\footnote{We disregard phenomena involving the whole Moon e.g., see Spinrad
1964, Sanduleak \& Stocke 1965, Verani et al.\ 2001, and events tied to solar
eclipses, or poorly localized.}

Several kinds of experiments on Apollo lunar missions, both orbiting and
on the surface, as well as on {\it Lunar Prospector}, were designed to detect
and identify gasses in the tenuous lunar atmosphere, both ions and neutral
species, plus decay products from associated gaseous radioactive isotopes.
Even though some of these spent only days or weeks operating near the Moon,
most observed evidence of sporadic outgassing activity, including events that
seem unassociated with anthropogenic effects.
(We analyze these in Crotts 2007a.)

On the timescale of a decade, numerous spacecraft and humans will visit the
Moon again.
This offers an unprecedented opportunity to study the atmosphere of the Moon,
but will also introduce transients from human activity that may complicate our
understanding of this gas and what it can disclose regarding the lunar
interior's structure, composition and evolution.
We must evaluate the current results now and expand upon them rapidly to
exploit our upcoming opportunity to explore the Moon in its still pristine
state and perhaps even exploit these still poorly understood resources.
We would like to evaluate if TLPs might be elevated into a tool which can be
used to study other events on the Moon, including outgassing, and thereby do
so from Earth, before The Return to The Moon of large spacecraft and their
human crews.
If TLPs are real, we can study them using modern technology without depending
on human event selection, but via a robotic imaging monitor that will be much
more objective and probably more sensitive than historical means (Crotts 2007b,
Crotts et al.\ 2007).

\section{Transient Lunar Phenomena}

\subsection{The Troublesome Nature of TLP Observations}

With the sensitivity of the human eye peering through an optical telescope, the
detection of a TLP is evidently a rare event.
Heretofore, this has put TLP reporting largely into the category of anecdotal
evidence, which in many minds makes them ``irreproducible.''~
The harshest critics of the field have likened them to UFOs (Unidentified
Flying Objects).
This is not to say that rare sightings and anecdotal evidence cannot turn into
real phenomena and important science e.g., meteorites, ball lightning, the
green flash, and many species of rare and interesting fauna.

The debate as to the reality of TLPs unfortunately has taken place in the
unrefereed scientific literature.
Additionally, in recent years there have been examples of published, apparently
positive evidence that has been later retracted.
A positive treatment is found in Cameron (1991).
Sheehan \& Dobbins (1999) present a condemning case.
Also see Haas (2003), and Lena \& Cook (2004).
In this paper we intend to give a fair evaluation of this situation, as
quantitatively as possible given the state of the data.

Some systematic searches have heretofore yielded few reliable detections; we
will raise the question whether these surveys were sufficiently comprehensive
to have produced a non-null result.
The power of the heterogeneous sample is the much greater coverage over years
and centuries compared to {\it in situ} spacecraft studies or even programmed
Earth-based, telescopic surveys.
For now we will consider the nature of the bulk of these reports, and evaluate
their utility in understanding physical phenomena.

One must assume one of the following: either TLPs are in part tied to physical
processes local to the vicinity of the lunar surface, or they are ``false,''
either due to human misinterpretation of normal lunar appearance due to
physical effects tied to terrestrial phenomena or even the delusion or
fabrication of the observer.
Given the complexity of the data base we are considering, one might even
consider a combination of all three.
It is the primary task of this work to answer the question if TLPs are at least
consistently tied to specific sites on the lunar surface, so that one can then
ask if the appearance or physics of these sites might explain TLP reports.

\subsection{The TLP Observers}

The compilation and cataloging of TLP reports is due largely to the massive
efforts of two dedicated investigators, Winifred Cameron (1978, with 1463
events through May 1977), and Barbara Middlehurst (1977a, primary from the TLP
event catalog of Middlehurst, Burley, Moore \& Welther 1968, of 579 events up
to 1968 October, with another 134 added by Patrick Moore through 1971 May).
All but seven of Cameron's were seen after the invention of the astronomical
telescope, and virtually all reports after the year 1610 were telescopic (at
least 1446 of the 1456, not counting several naked-eye observations by Apollo
astronauts from lunar orbit).

Most of the naked-eye reports (without telescope) describe bright spots on the
daylit or darkside Moon, often seen by several observers, and these seem
possibly consistent with particularly bright examples of the kinds of spots
seen in profusion with the aid of telescopes.
None of these events was reported by observers in widely-separated locations,
and seem to be recorded only a few times or less per century.
(An example: Boston, Massachusetts, the evening of 1668 November 26; several
naked-eye observers report a bright, star-like point on the dark side: 
Middlehurst 1977a, from Jocelyn 1675.)~
An exceptional case is the report from 1178 June 18 by at least five observers
in Canterbury, England (Newton 1972, Stubbs 1879), an event which Hartung
(1976) speculates might be the impact formation of the young crater Giordano
Bruno, as supported by Calame \& Mulholland (1978) searching for a source of
the anomalous lunar libration, but seriously challenged by Gault \& Schultz
(1991), Withers (2001), and unpublished work.

The overwhelming majority of TLP reports involve telescopic observations, and
most of these were made by amateur astronomers.
During the years of the first lunar surface exploration efforts, many sightings
were made by mixed teams of professionals and amateurs, detailed in \S 2.4.
At times when the Moon has been an attractive target of forefront research,
some of the most noted and experienced observers have reported TLPs.
(Middlehurst 1977a gives a separate summary of this early period of TLP
reports.)~
Wilhelm Herschel, the only human to discover a planet on the basis of solely
visual telescopic observation, reported TLPs on at least 6 occasions between
1783 and 1790, primarily bright, point-like spots, many of them red in color
(Middlehurst 1968).
Clyde Tombaugh, discoverer of the dwarf planet 134340 Pluto, along with at
least five other observers spread across the United States, reported on 1963
November 28 ``reddish-orange and sparkle'' activity on the rim of Aristarchus,
which some observers saw followed by a faint blue glow on the crater floor
(Cameron 1978).
Charles Messier, discoverer of 19 comets and author of the famous catalog of
nebulae, saw TLPs on one occasion (``moving glows'' during a lunar eclipse:
Cameron 1978) in 1783, and Ernst Tempel (discoverer of 20 or more comets) at
four epochs from 1866-1885.
TLPs were also reported by noted observers Edward Barnard (in 1889-1892),
Edmond Halley (in 1715), Johann Bode (in 1788-1792), George Airy (in 1877, and
confirmed independently), Heinrich Olbers (in 1821), Dominique Cassini in
1671-1673 (Paris Observatory director and grandson of Jean-Domenique Cassini),
Camille Flammarion (in 1867-1906), William Pickering (perhaps the first to
place a tight astrophysical limit on the Moon's atmospheric mass) in 1891-1912,
Johann Schr\"oter in 1784-1792 (first to notice the phase anomaly of Venus),
Friedrich von Struve (in 1822), Francesco Bianchini (in 1685-1725), Etienne
Trouvelot (in 1870-1877), and more recently Zden\v ek Kopal (in 1963) and in
1948-1967, Sir Patrick Moore (Middlehurst 1977a, Cameron 1978).
Of course, in 1821-1839 Franz von Gruithuisen reported luminous and obscured
spots on the Moon, yet also wrote about the Moon being inhabited and dotted by
cities!
(He was also first to conclude that lunar craters result from meteorite
impacts.)

Several reports by simultaneous but geographically well-separated observers of
the same events on the lunar surface are recorded e.g., 1895 May 2, for
12-14 min on the floor of crater Plato, Brenner reported a streak of light,
while (independently?) Fauth reported bright, parallel bands.
Cameron (1991) describes the observations by Greenacre and Barr on 1963 October
30 of several reddish spots that appeared for several minutes in Aristarchus
and near Schr\"oter's Valley, and were also seen by other observers.
Similar events occurred one month later (see above) in the same vicinity; both
cases roughly coincident with local sunrise.
Apollo astronauts Cernan, Schmidt, Mattingly, Aldrin, Collins (and Armstrong?)
all reported TLPs from lunar orbit, on four occasions.
Three of these were rapid flashes that have been hypothesized to result from
cosmic rays entering their visual system, but on {\it Apollo 11}, Aldrin and
Collins reported a strange darkside surface appearance (Jones 1969) during a
1-2 minute period in which ground-based observers saw a similar phenomenon at
likely the same location (Cameron 1978).
We discuss this singular case in detail in Appendix I.

\subsection{Photographic Evidence}

There are at least nine events noted by Cameron (1978) as having been
photographed, the earliest from 1953 November 15.
Many of these are unpublished, but there are some dramatic exceptions.
On 1956 October 26, D.\ Alter (1957) took a careful sequence of photographs on
Mt.\ Wilson Observatory 60-inch of the craters Alphonsus and Arzachel, in 
infrared (Kodak I-N emulsion) and blue-violet (II-O) light, which allow a
differential measurement of the imaging properties in time and wavelength
between the two craters.
There is a perhaps convincing, apparent obscuration of the floor of Alphonsus
not seen later in time or in Arzachel, and that is apparent in the violet but
not infrared as if some scattering cloud is present.
A similar effect in crater Purbach was photographed on 1970 April 14 by Osawa
but not published (Cameron 1978).
On 1959 January 23, Alter recorded (but never published) a photograph of a
bright blue glow on the Aristarchus floor, which then turned white.
Two unpublished event photographs (\#876 and \#1145 in Cameron 1978) are
claimed to show red spots in craters Aristarchus and Maskelyne, respectively,
with the former event apparently confirmed by separate visual observers.
Two other unpublished photographs involve brightenings of Aristarchus.
More recently Cameron (1991) presents fairly dramatic photographs of a glowing, 
reddish-gray patch moving on the floor of crater Piticus, as observed by
G.\ Slayton on 1981 September 5.
Finally, during a polarimetric program at l'Observatoire de Paris for lunar
surface texture analysis, Dollfus (2000) caught on 1992 December 30 a
brightening in the center of crater Langrenus, and with it an associated
increase in the degree of polarization.
Similar polarimetric changes were recorded on at least two occasions in
Aristarchus, but the timescale is unclear (Dzhapiashvili \& Ksanfomaliti 1962,
Lipsky \& Pospergelis 1966).
We will discuss spectroscopic and polarimetric observations in Paper II, and
several probable meteoritic impact photographs in \S 2.5 below.


\subsection{Patrols and Systematic TLP Searches}

Several programs, primarily by groups of amateur astronomers but sometimes
involving professional researchers, have made organized observations of the
Moon with the goal of constructing scientifically more useful datasets in
attacking the TLP problem.
Several of these have been organized by D.\ Darling and collaborators, often in
connection with ALPO (Association of Lunar and Planetary Observers:
http://labbey.com/ALPO/Lunar.html and
http://www.lpl.arizona.edu/rhill/alpo/lunar.html) and now with the BAA
(British Astronomical Association: http://www.britastro.org/baa/ and
http://www.cs.nott.ac.uk/acc/).
A summary of activities until fairly recently is found at
http://www.ltpresearch.org/.
(Other groups can also be found at http://www.glrgroup.org/.)~
An informal appraisal of the information from these groups indicates that
observers in these programs are patrolling for TLPs at the level of order
100 h per year.
A summary follows of some of these past efforts bearing on the purposes of this
paper.

\noindent
{\bf Operation Moon Blink} (August 1964 - April 1966): W.\ Cameron (1966;
Cameron \& Gilheany 1967) of NASA Goddard SFC organized a network of
professional and amateur observatories spanning the contiguous United States
and provided 12 of them with an instrument designed to be particularly
sensitive to transient sources, particularly anomalously-colored ones, on the
lunar surface.
Several more observers were engaged in an alert follow-up network using more
conventional visual and photographic techniques.
The Moon-Blink apparatus consisted of an S-20 photocathode image tube with a
rotating filter wheel cycling between a red filter (similar to Wratten 29) and
blue (like Wratten 44a) at an adjustable rate of 4-12 Hz.
The video output was then monitored visually for any blinking sources.
This device was particularly sensitive to color changes, shown by tests to be
limited at the 0.02 mag level to changes in color index, but relatively
insensitive to color-neutral brightenings, dimmings and obscurations.
During the survey 25 events were reported, 14 of which were detected as color
changes.
No total observing time estimate was published, but of order 1000-2000 hours
was devoted to the blink portion (Cameron 2006).
Eight of these were actual color blink events.
Two of these (one of the blink events) were confirmed by other means, but most
events were reported by a team of observers at a single site.
Observations were concentrated on small areas, particularly Aristarchus, and
the fields of view varied with a selection of different telescopes above
15-inch diameter.
The investigation produced no strong conclusions regarding TLPs.

\noindent
{\bf Corralitos Observatory L.T.P.\ Monitoring Program} (October 1965 - April
1972):  J.A.\ Hynek et al.\ (1976; Dunlap \& Hynek 1973, Hynek \& Dunlap 1968)
conducted a monitoring program for TLPs using a 62 cm telescope and an S-20
image tube cycling through three bands spanning the optical spectrum in a
``blink'' mode, and logged 3000 h of observation by 1968 and 8000 h by 1972.
The field-of-view was 0.1 degree.
(There is a reference to ``man-hours'' of observation, so perhaps the time
coverage is two or three times less than this.)~
During this time, there was only one transient event reported in their
publications, a large-scale violet excess just prior to the lunar eclipse of
1967 April 24, not even classified as a TLP for the purposes of this paper.
Several TLP alerts by outside observers were transmitted to Corralitos
Observatory during the events, and apparently most of these were negative
confirmations.
Cameron (1978) lists 25 events that were apparently negative (four where this
is stated explicitly in terms of the data), and two (\#1119 and \#1150) where
apparently Cameron disagrees with Hynek et al.\ and concludes there was a
positive confirmation.
Dunlap \& Hynek (1973) claim a sensitivity limit of better than 5\% change in
intensity in a 100\AA~band.
Hynek et al.\ speculate that some TLPs are actually due to observers'
misinterpretation of rapid seeing in changes.

\noindent
{\bf Lunar International Observing Network} (during Apollo missions 8 \& 10-12,
1968-1969):
B.\ Middlehurst (1970) organized LION, consisting of 216 observing stations in
30 countries effectively covering all longitudes.
LION performed systematic observations at preselected times of lunar features
with a history of TLP reports.
This produced 169 reports for 31 lunar areas from 28 observing stations located
in 19 countries, during Apollo missions 10-12.
In particular a special campaign produced the confirmation including the {\it
Apollo 11} event described above.
This one effort involved the crater Aristarchus on 19 July 1969, from 18:45 to
24:00 U.T.
Twelve observers in six countries and two continents made simultaneous or
overlapping observations of the Aristarchus crater covering a 5.25 h time
interval.

\noindent
{\bf ALPO Clementine Campaign} (1994 February 19 - May 3):
The Association of Lunar and Planetary Observers (Darling 2006) organized
47 observers during the 71-day Clementine multispectral mapping mission,
producing five probable event detections (according to a weighting scheme
developed by W.\ Cameron), some seen by over 40 observers at a time.
There are four instances in which Clementine multispectral images were acquired
both before and after one of these reports (Buratti et al.\ 2000).
Despite initial indications to the contrary (Buratti et al.\ 1999), none of
these four sets of images shows clear changes (that would be of a
semi-permanent nature) that could be attributed to these TLPs.

\subsection{Description and Distribution of TLP Reports}

As we have said, Cameron (1972) splits TLPs into brightenings, red and
blue-colored events, and dimmings plus obscurations.
Of 113 reports in Middlehurst (1977a) involving enhanced brightness in blue
and/or violet, 101 of them involve J.C.\ Bartlett, composing most of his total
of 114 reports (between 1949 and 1966), most of those (108) involving
Aristarchus.
In contrast only 9 of 12 total non-Bartlett blue/violet events occur in the
same years (during which 47\% of all reports occur).
We must correct for this somehow, either by rejecting all blue/violet events or
all reports by Bartlett; we choose the latter.

\subsubsection{Timescale Distribution}

Seventy-one reports in Middlehurst et al.\ (1968) include duration estimates
interpretable to better than a factor of two.
This is not a statistical sample, but give some measure of event duration;
binned in $\sqrt{10}$ intervals from 60s to 19000s (with the
longest event being 18000s and the shortest 60s) the duration distribution is:
60-190s, 7 reports; 191-600s: 9; 601-1900s, 27, 1901-6000s, 23; more than
6000s: 5.
These effects are sufficiently prolonged to allow reinspection (albeit by the
same observer in most cases).
Nonetheless, during the observations, internal changes are often seen on rapid
timescales (selected from Cameron 1978, Middlehurst et al.\ 1968: ``Abrupt
flash of red settling immediately to point of red haze,''  ``A series of weak
glows; Final flash observed at 04h18m,'' ``White obscuration moved 20 mph, 
decreased in extent.  Phenomenon repeated,'' etc.).

If TLPs are caused by impacts, they are caused by phenomena at the lunar
surface, but will be largely uncorrelated with lunar location.
There are four cases in Middlehurst et al.\ (1968) described as sudden,
isolated flashes of light, and these are not correlated with meteor showers
(occurring on 1945 October 19, 1955 April 24, 1957 October 12, and 1967
September 11).
None of these are well-placed with respect to known meteor showers.
(April 23 is the peak of the Pi Puppids, but these are strong only near the
perihelion of comet 26P/Grigg-Skjellerup, which occurred in 1952 and 1957, not
1955.)~ 
Suggestions for other mechanisms for short-lived TLPs include piezoelectric
discharge (Kolovos et al.\ 1988, 1992 - which also includes an interesting
recorded TLP observation).

How does a meteorite impact appear on the surface of the Moon?
Several of these events have probably been observed recently (since the Cameron
and Middlehurst TLP catalogs), although some have not yet been published.
(See Cudnik et al.\ 2003 for some interesting discussion, Cudnik et al.\ 2007
for a recent summary, and Hunton et al.\ 1991 for another lunar impact detection
idea.)~
Five Leonid events were reported by Ortiz et al.\ (2000), three of them
confirmed by simultaneous observers.
A patrol using a double-detector coincidence system detected three probable
meteorite hits (Anoshkin, Petrov \& Mench 1978, also see Arkhipov 1991), and
events have been caught by Dunlop (1999) and Suggs et al.\ (2005, 2006).
While the latter have not been published, the available data show that 
meteorite impacts tend to be rapid, with an exponential decay times of about
0.1s.
Other works include Ortiz et al.\ 2007, Volvach et al.\ 2005, Chandrasekhar et
al.\ 2003 and Cooke et al.\ 2007.
Multiple observer confirmation of a small Perseid meteorite impact (Yanagisawa
et al.\ 2006) indicates a timescale about three times shorter, whereas the
possible photographic record (Stuart 1956) of a large impact on 1953 November
15 (near 5$^\circ$N, 3$^\circ$W on the Moon) lasting at most 8 seconds
(emitting electromagnetically $3\times 10^{18}$ erg s$^{-1}$ and perhaps as
massive as $10^{13}$ g: Buratti et al.\ 2003) was at one time thought to have
been confirmed by the coincidence of a fresh crater seen by {\it Clementine}
(Buratti et al.\ 2003) but was contradicted by pre-event photographs (Beatty
2003).
Some impacts might involve space debris (see Maley 1991, Rast 1991).

We have one unquestionable detection of a ``meteorite'' hit on the Moon in the
form of the spacecraft {\it SMART-1}'s impact on 2006 September 3.
{\it SMART-1} at the time had a total\footnote{Foing, B.\ (2006), personal
communication, although the dry mass of {\it SMART-1} is listed as 305 kg:
http://nssdc.gsfc.nasa.gov/database/MasterCatalog?sc=2003-043C}
mass of 280 kg and was moving 2 km s$^{-1}$.
This is the kinetic energy of a typical meteoroid measured in the lunar frame
(30--40 km s$^{-1}$) of about 1 kg, several times smaller than what models
indicate was required for the Suggs et al.\ events above.
Indeed, an array of telescopes with optical CCD imagers and diameters up to 1 m
observed the impact with negative results (Ehrenfreund 2006).
In the near IR, however, the situation was much different: as yet unpublished
results (Veillet 2006) from a 10~s exposure using the WIRCam infrared imager on
the 3.6~m Canada-France-Hawaii Telescope at the time of impact show a signal so
bright that it saturated the detector in a 32 nm band at the 2.122~$\mu$m
molecular H$_2$ S(1) 1$\rightarrow$0 transition.
The signal detected was at least $3 \times 10^6$ e$^-$ and probably many times
more, which corresponds to at least $8\times 10^{-15}$ erg cm$^{-2}$.
From stars in the field beyond the Moon at the same time, one can estimate the
seeing at about 1.5 arcsec FWHM, indicating that the flux is probably about
5 times higher at least (which cannot be fully estimated without careful
non-linearity tests or models), meaning that the energy output in this band was
at least about $10^9$ erg, about $2\times 10^{-7}$ of the total
energy and probably two orders of magnitude more than the limits that will be
derived from the optical non-detections, if as reported.
There is also a luminous debris cloud spreading elongatedly starting at
$\sim$1 km s$^{-1}$ (even though the bright impact source was nearly
point-like), as evident on the subsequent images, and which likely carried
a large fraction of the energy and largely disappeared over 150 s.


The {\it SMART-1} impact was very atypical of a meteoroid impact, since not
only was it much slower but impacted at an angle of only $3^\circ$ with respect
to the horizon.
Furthermore, the impactor consisted not only of spacecraft structure but
significant amounts of hydrazine, which probably broke down immediately into
atomic and molecular N and H (and perhaps NH$_3$) and charged states thereof,
possibly adding considerably to the specific wavelength band chosen for the
CFHT detection (but also possibly to Balmer and Paschen lines in the optical
and near IR accessible to Si CCDs, and even compounds with regolith material,
predominantly O, e.g., near IR/optical Meinel bands of OH).
Note as well that the impact at 1.7 km s$^{-1}$ of the 158-kg {\it Lunar
Prospector} into a permanently shadowed polar crater produced no unambiguous
detection in the optical or radio (Barker et al.\ 1999, Berezhnoy et
al.\ 2000).
These suggest that further studies of meteoritic impacts on the Moon might
benefit from use of an IR camera system.

This is supported by the recent presentation (Svedhem 2006) showing similar
results for the impact of the spacecraft $Hiten$ near crater Stevinus on the
daytime nearside highlands.
The spacecraft of mass 143 kg struck at 2.323.km s$^{-1}$ at 48$^\circ$ from
vertical, with kinetic energy of $3.9 \times 10^{15}$ erg.
D.~Allen used the AAT to observe the impact at 2.16~$\mu$m wavelength and
recorded an event fluence corresponding to $6.7 \times 10^{12}$ erg in a 1\%
wavelength band, appearing 6-16 s after impact
No immediate optical flash was seen and later optical signals (15 min after
impact) were unclear, and not near the original impact point.
Svedhem likewise concludes that emission from the 1 kg of hydrazine onboard is
likely an important part of the infrared source; this may be a key difference
between these spacecraft impacts and meteoritic impacts in the infrared.

Regardless of the details of the above, it is clear that the great majority of
TLP reports are not impact events.
Even if very large impacts can produce events of sufficiently long duration, it
is clear from model computation e.g., Morrison et al.\ (1993) that the fresh
impacts seen in {\it Clementine} and other data sets cannot sustain such
activity.

\subsubsection{Spatial Distribution}

Since meteoritic impact cannot be the cause of transient on the timescales seen
in the great majority of TLPs, we can expect that the spatial distribution
might be expected to carry detailed information about the TLP mechanisms (if
observer selection effects can be removed).
How are TLPs localized on the lunar surface?
Table 1 is derived from reports listed by Middlehurst et al.\ (1968), sometimes
with additional information (but not additional reports) drawn from Cameron
(1978).
This information is summarized in Figure 1.

\bigskip
\noindent
Table 1: Number of TLPs Reported, by Feature
\begin{verbatim}
_______________________________________________________________________________
Raw
Report   Feature (Latitude, Longitude)
Count
------   -----------------------------
122      Aristarchus (24N 48W)
 40      Plato (51N 09W)
 20      Schroter's Valley (26N 52W)
 18      Alphonsus (13S 03W)
 16      Gassendi (18S 40W)
 13      Ross D (12N 22E)
 12      Mare Crisium (18N 58E)

6 each   Cobra Head (24N 48W); Copernicus (10N 20W); Kepler (08N 38W);
         Posidonius (32N 30E); Tycho (43S 11W)

5 each   Eratosthenes (15N 11W); Messier (02N 48E)

4 each   Grimaldi (06S 68W); Lichtenberg (32N 68W); Mons Piton (41N 01W);
         Picard (15N 55E) 

3 each   Capuanus (34S 26W); Cassini (40N 05E); Eudoxus (44N 16E);
         Mons Pico (B) (46N 09W); Pitatus (30S 13W); Proclus (16N 47E);
         Ptolemaeus (09S 02W); Riccioli (03S 74W); Schickard (44S 26E);
         Theophilus (12S 26E)

2 each   1.3' S.E. of Plato (47N 03W); Alpetragius (16S 05W); Atlas (47N 44E);
         Bessel (22N 18E); Calippus (39N 11E); Helicon (40N 23W);
         Herodotus (23N 50W); Littrow (21N 31E); Macrobius (21N 46E);
         Mare Humorum (24S 39W); Mare Tranquilitaties (08N 28E);
         Mons La Hire (28N 26W); Montes Alps, S. of (46N 02E);
         Montes Teneriffe (47N 13W); Pallas (05N 02W);
         Promontorium Agarum (18N 58E); Promontorium Heraclides (14N 66E);
         South Pole (90S 00E); Theaetetus (37N 06E); Timocharis (27N 13W)

1 each   Agrippa (04N 11E); Anaximander (67N 51W); Archimedes (30N 04W);
         Arzachel (18S 02W); Birt (22S 09W); Carlini (34N 24W);
         Cavendish (24S 54W); Censorinus (00N 32E); Clavius (58S 14W);
         Conon (22N 02E); Daniell (35N 31E); Darwin (20S 69W); Dawes (17N 26E);
         Dionysius (03N 17E); Endymion (54N 56E); Fracastorius (21S 33E);
         Godin (02N 10E); Hansteen (11S 52W); Hercules (47N 39E);
         Herschel (06S 02W); Humboldt (27S 80E); Hyginus N (08N 06E);
         Kant (11S 20E); Kunowsky (03N 32W); Lambert (26N 21W);
         Langrenus (09S 61E); Leibnitz Mt. (unoffic.: 83S 39W);
         Manilius (15N 09E); Mare Humorum (24S 39W); Mare Nubium (10S 15W);
         Mare Serenitatis (28N 18E); Mare Vaporum (13N 03E); Marius (12N 51W);
         Menelaus (16N 16E); Mersenius (22S 49W); Mont Blanc (45N 00E);
         Montes Carpatus (15N 25W); Montes Taurus (26N 36E); Peirce A (18N 53E);
         Philolaus (72N 32W); Plinius (15N 24E); Sabine (01N 20E);
         Sinus Iridum, S. of (45N 32W); Sulpicius Gallus (20N 12E);
         Taruntius (06N 46E); Thales (62N 50E); Triesnecker (04N 04E);
         Vitruvius (18N 31E); Walter (33S 00E); 
                                    
Not counted: 4 (global lunar changes), 14 ("cusp" events), 43 (events w/
unknown coordinates)
_______________________________________________________________________________
\end{verbatim}

The spatial modulation of the report rate, beyond just the frequency at
specific sites, is dominated by the tendency of reports to avoid the deep
highlands and to some degree the mid-mare plains, congregating instead near the
maria/highland boundary (Cameron 1967, 1972, Middlehurst \& Moore 1967,
Buratti et al.\ 2000).
Even Aristarchus in the midst of Oceanus Procellarum rests on a giant block
(probably from a previous mare basin impact) elevated 2 km above the mare plain.
TLP reports favor the western half of the near side (106 in the east longitude,
166 in the west not counting an additional 144 in the west on the Aristarchus
Plateau), counter to the usual preference of casual observers to observe
earlier in the night, perhaps due to the greater extent of maria (and maria
boundaries) on the western side.
We will return to this discussion after we deal with at least some aspects of
observer selection effects.

\subsection{Observer Selection Bias and Correlation Effects}

To make further progress in understanding the spatial distribution of TLPs, we
must deal statistically with the horrendous selection effects introduced into
these data by the patterns and biases of the observers, most of whom never
intended that their reports form part of a statistical database.
Our task is not modest; we are basically trying to calibrate for this purpose
all of the observations made (or not made) of the Moon by all of the human
eyeballs over a period of centuries.
How do we possibly deal with the historical and even psychological issues that
these effect imply, as well as the physical/mathematical ones?
This is the major burden of the current paper; nonetheless, there are some
regularities that we might exploit.

There are many works on selenography, but in general, there were some
systematic naked-eye observations of the Moon starting in Europe in the
1400--1500's, and much earlier in China (but I find no early chinese reports of
TLPs).~
Observations greatly increased in number and detail in the mid 17th century
after the invention of the telescope, and early in the next century it was
appreciated that the Moon needed to be observed more often due to the effects
of libration.
Still, it does not seem that attention focussed solely on the limb regions, as
mapping the Moon increased in detail generally, with the adoption of lunar
coordinates by mid-to-late 18th century.
From late-1700's to early 1900's, visual mapping of the entire Moon was an
active science, perhaps with some concentration on terminator regions in order
to better sense relative elevations of lunar features.
By the turn of the 20th century, visual observation was increasingly replaced
by photography, which had the unfortunate effect of suppressing sensitivity to
TLPs due to the decreased observational sampling cadence of photographic
plates with respect to the human eye, as well as the loss of prompt color
information.
I speculate that for this reason, reports of TLPs by professional astronomers
began to die out.
It is not the purpose of this paper to dwell on the history of TLP observations
beyond gaining some insight into the statistical sampling structure and false
report rate of the catalogs.
Shortly, we will move on to questions more objectively dependent on the physics
of the lunar surface.

There is a pause in the frequency in TLP reports in both the Cameron (1978) and
Middlehurst (1968) catalogs, and indeed the break in reports 1927-1931 divides
the Middlehurst catalog at the median epoch in the catalog.
For the post-1930 half of the sample, 2/3 of the reports come after 1955, at
which time TLPs become much more commonly known and observing pattern appear to
change, as we will describe later.
For now we will concentrate awhile on the pre-1956 and particularly the
pre-1930 period.

It is obvious in regard to TLPs that a most unusual area is the Aristarchus
Plateau, including the crater Aristarchus itself, the adjacent crater
Herodotus, and Vallis Schr\"oteri (Schr\"oter's Valley) flowing from ``Cobra's
Head'' (``Cobra-Head''), together occupying the southeasternmost $\sim$10,000
km$^2$ of the $\sim$50,000 km$^2$ Plateau in the midst of the huge ($4\times
10^6$ km$^2$) mare region Oceanus Procellarum.
(Vallis Schr\"oteri was once selected as the landing site for Apollo 18, later
cancelled along with Apollos 19 and 20.)~
Aristarchus is among the brightest lunar craters, sometimes {\it the}
brightest, sometimes visible to the unaided eye from Earth, as noted as far in
the past as the Tang Dynasty (618 -- 907 A.D./CE) (Mayers 1874).
It is also one of the freshest: $\sim 500$ My old, along with Copernicus,
Kepler and Tycho (each producing less than 5\% of the TLP reports of
Aristarchus.
At one time the region was intensely active with volcanic flows and eruptions,
and many sinuous rilles remain, likely old lava channels, including the most
voluminous on the Moon, Schr\"oter's Valley.

Even more than Copernicus, Aristarchus is singular in standing in such contrast
to the surrounding dark mare background, although this is not true of Vallis
Schr\"oteri/Cobra's Head or Herodotus also on the Plateau.
The Aristarchus region is responsible for $\sim$50\% of the visual TLP reports
(but also likely receives a large fraction of the observing attention).
As reviewed in (Crotts 2007a), however, the Aristarchus Plateau is also
responsible for undeniably objective lunar anomalies of a transient nature
associated with lunar outgassing.
We will use Aristarchus as a proxy to trace how astronomers have observed TLPs,
and whether these observations have influenced each other during different
historical periods thereby producing correlated observations, rather than
reports that can be counted individually.

I cannot presume to appreciate the observing motivations of astronomers from
centuries, but prior to the year 1956 there appears to be little writing
labeling special sites such as Aristarchus as targets of particular popular or
professional attention in terms of TLPs.
Aristarchus received closer scrutiny in 1911 with R.\ Wood indicating that it
might contain high concentrations of sulfur, but this did not produce a spate
of Aristarchus TLP reports.
Indeed, Wood discusses vulcanism in the context of Aristarchus (sometimes known
as ``Wood's Spot''\footnote{e.g., Whitaker (1972), or
http://www.lpod.org/archive/archive/2004/01/LPOD-2004-01-17.htm}) and seems
unaware of the number of TLP reports in the vicinity (Wood 1911).
Furthermore, Birt (1870) and Whitley (1870) provide a historical overview
(1787--1880) of visual observations of Aristarchus (and Herodotus) while
conducting a spirited debate about the nature of features including possible
changes in their appearance.
They mention small, possible changes, but give them no special
significance, nor mention anything that today we might refer to as a
recognized TLP phenomenon (or at least a human tendency to report TLPs).
A different statement is made by Elgers (1884), who again reviews Aristarchus,
Herodotus and the surrounding plateau.
While he does not mention anything like TLPs, he makes a telling statement:
``Although no part of the moon's visible surface has been more frequently
scrutinized by observers than the rugged and very interesting region which
includes these these beautiful objects, selenographers can only give an
incomplete and unsatisfactory account of it...''
By 1913, however, there appears to be some scholarly awareness of reported
activity at Aristarchus; witness the summary (Maunder 1913) of reports by
Herschel that he ``thought he was watching a lunar volcano in eruption'', and
by Molesworth and Goodacre who ``each on more than one occasion, observed what
seemed to be a faint bluish mist on the inner slope of the east wall... for a
short time.
Other selenographers too, on rare occasions, have made observations accordant
with these, relating to various regions on the Moon.''
Maunder balances this with skepticism e.g., ``one of the most industrious of
the present-day observers of the Moon, M.\ Philip Fauth, declares that as a
student of the Moon for the last twenty years, and as probably one of the few
living investigators who have kept in practical touch with the results of
selenography, he is bound to express his conviction that no eye has ever seen a
physical change in the plastic features of the Moon's surface'' (citing Fauth
1909.\footnote{This skepticism of Fauth's was despite his defense in 1912, to
his scientific disgrace, of the Austrian engineer Hanns H\"orbiger's curious
notion that the Moon's ``ice-like'' appearance implies that it and other
celestial objects must be composed of ice! (H\"orbiger \& Fauth 1913)})

The periods covered by these papers are particularly telling in understanding
the nature of human behavior over the changing technological state-of-the-art
in the historical course of selenography.
A paper by W.H.\ Pickering (1892) asks ``Are there present Active Volcanos upon
the Moon?'' and does a quantitative study of candidate volcanoes on the Moon,
merging two lists with a total of 67 craters, and then discussing in turn many
of the 32 craters common to both lists.
Most of these are then eliminated for various reasons, then he starts
discussing the rest in turn.
While Aristarchus is mentioned, it is discounted as being a non-volcano, while
several other features are taken much more seriously (Bessel, Linn\'e and
Plato), and there is {\it no} discussion of any activity that we would call
TLPs today, only the long-timescale appearance or permanent changes.
Pickering described several TLP reports shortly before and then after
submitting this paper --- statistically marginal in themselves, 14 in our
sample, and 9\% of Aristarchus and vicinity --- but I see no significant
evidence that these induced a spate of further Aristarchus reports, at least
until publication of his book [1904a], and probably not even then.
This was rather late in the pre-1930 period, anyway (which we investigate more
quantitatively below).

As is the case for the whole sample, we should search the Aristarchus reports
for observers e.g., Bartlett, who produce obvious statistically descrepant
results.
With 150 reports total for Aristarchus, Pickering should be considered as a
marginal candidate for this.
His series of 14 reports from 1891-1898 refer to mists or nebulosity in the
Herodotus/Cobra Head vicinity east and north of Aristarchus (Pickering 1900).
Personally, I suspect that at least some of these reports were erroneous, due
Pickering's tendency to overinterpret observations according to his lunar
world view.
If these were inconsistent with other observers' body of reports and the
catalog as a whole, and might be considered a candidate for exclusion.
In truth there are contemporaneous reports e.g., Molesworth and Goodacre in
1895-1897 (Goodacre 1899, 1931) describing mists and darkening nebulosities
around Aristarchus.
Goodacre and Molesworth were based in Britain and what was then Ceylon, and it
is not apparent that they were influenced by Pickering.
Strictly speaking we cannot exclude the Pickering observations because of an
inconsistency.
Pickering exhibited a tendency to interpret changes in the lunar surface in
terms of weather and biology, but his statements regarding these issues
indicate that his statements are not motivated ideologically, but by the best
description of observation e.g., Pickering (1904b, 1916).\footnote{From
Pickering (1916): ``The writer has sometimes
been asked, `What reason is there to believe that there is ice upon the moon?'
The answer is: `For the same reason that we believe there is ice upon Mars,
because the phenomena observed can be more readily explained that way than any
other.'"
From Pickering (1904b), commenting on his interpretation of mists over the
surface of the Moon, which he had established by occultation studies to be at
very low atmospheric pressure: ``It seems to the writer that the merit of this
explanation lies not so much in its novelty, but rather because it is founded
so largely upon the observed facts.''
W.H.\ Pickering is an interesting case!
}
In the end the Pickering subsample has little effect on the overall qualitative
behavior of the Aristarchus dataset anyway, so I leave it intact.

To consider the questions above more systematically, I make use of
abstract/article search engines.
A search of the Astrophysics Data System (ADS)\footnote{The
Smithsonian/NASA Astrophysics Data System:
http://www.adsabs.harvard.edu/ - which covers a great many journals into the
19th century and even earlier e.g., Lind \& Maskelyne 1769, or Street et
al.\ 1671} Astronomy  and Physics archive before 1930, in the title or
abstract, for ``volcano'' and (not or) ``aristarchus'' produces no matches,
whereas ``aristarchus'' alone produces 4 matches (not counting the ancient
greek astronomer Aristarchus!) and ``volcano'' alone produces 13 relevant to
the Moon.
Note that this search does not go into the body (versus the abstract) of the
text of some longer articles, for example Wood 1911, but perhaps this is a
satisfactory reflection of the amount of attention that a TLP-like claim might
attract.
Similarly results are found for ``gas,'' ``atmosphere,'' ``eruption,''
``flash,'' ``cloud,'' ``nebulosity,'' ``mist,'' ``geyser,'' and ``vapo(u)r,''
with no matches for any of these with ``aristarchus.''
These results are summarized in Table 2.
Likewise, replacing ``aristarchus'' with ``herodotus'' or various versions of
Schr\"oter's Valley or Cobra's Head produce no matches with the above terms for
potentially TLP-like phenomena, or any articles which described TLPs.

\noindent
Table 4: TLP-related Terms Correlated with Terms ``Aristarchus'' and ``Moon,''
before 1930
\begin{verbatim}
________________________________________________________
                            "moon/lunar"   "aristarchus"
                Number of      Cross-         Cross-
Search Term     Citations     Citations      Citations
_____________   _________   ____________   _____________

"volcano"           38           14              0
"gas"             1194            0              0
"atmosphere"       599           22              0
"eruption"          52            1              0
"flash"             89            0              0
"cloud"            192            3              0
"nebulosity"        45            0              0
"mist"               2            1              0
"geyser"             3            0              0
"vapo(u)r"         514            0              0
"transient"          9            0              0
"change"           893           38              0
No First Term     ----         2930              7
________________________________________________________
\end{verbatim}

I think that all of the above this is compelling evidence that the great
majority, perhaps all, pre-1930 selenographers did not place any special
importance on possible, specifically-localized TLP activity, particularly in
Aristarchus as examined here.\footnote{Let me say as a personal statement that
the degree of quantitative specificity and careful language by many of the
professional selenographers of these times is what made this investigation
possible.
In contrast, consider data like Bird's (1870) description of daytime meteors:
``they sail across the field of view like feather in the wind'' would be very
difficult to convert into a selection function or error rate.}

Again, we must search the 1930-1955 database for any observer correlation
effects visible in the literature, again using ADS Astronomy/Physics searches
on key terms (Table 3).
The only citation involving ``change'' AND ``aristarchus'' is Haas (1938)
referring to periodic changes appearance of the inner eastern wall of
Aristarchus over nine-day intervals, hence not TLPs in any regard.
The two other Aristarchus citations (Barcroft 1940, Barker 1942) concern the
same subject and were evidently written in reaction to Haas (1938).
This is the type of statistical correlation between events we would guard
against if these involved TLPs.

\noindent
Table 3: TLP-related Terms Correlated with Terms ``Aristarchus'' and
``Moon,'' 1930-1955
\begin{verbatim}
________________________________________________________
                            "moon/lunar"   "aristarchus"
                Number of      Cross-         Cross-
Search Term     Citations     Citations      Citations
_____________   _________   ____________   _____________

"volcano"           29            6              0
"gas"             2214            3              0
"atmosphere"      2234           47              0
"eruption"         119            2              0
"flash"             99            3              0
"cloud"           1036            9              0
"nebulosity"        81            1              0
"mist"               5            0              0
"geyser"             1            0              0
"vapo(u)r"         698            2              0
"transient"        112            1              0
"change"          2050           26              1
No First Term     ----         1124              3
________________________________________________________
\end{verbatim}

During 1930-1955, of the 26 citations involving ``moon'' AND ``change'' there
are six articles that actually deal with changes in lunar appearance, all on
time scales of days or weeks, spread over many lunar features, not
concentrating on any strong TLP sites (except for Haas [1938]).
These sites include Atlas, Billy, Cr\"uger, Endymion, Eratosthenes, Eudoxus,
Furnerius, Grimaldi, Hercules, Linn\'e, Macrobius, Mare Crisium, Messier,
Phocylides, Pickering, Pico/Pico B, Plato, Riccioli, Rocca, Snellius, Stevinus,
and Theophilus.
The only citation involving ``moon'' AND ``transient'' is irrelevant.

The date 1956 is significant because it delineates the period after which large
number of TLP reports appeared, starting with Alter (1957) and followed soon
thereafter by publications of Kozyrev (1959, 1962), inspired further
observations in a cascade through the catalog.
This wreaks havoc with our ability to evaluate TLP observational biases.
Citation correlation data for the period after 1955 is shown in Table 4.

\noindent
Table 4: TLP-related Terms Correlated with References to ``Aristarchus'' and
``Moon'', 1956-1968
\begin{verbatim}
________________________________________________________
                            "moon/lunar"   "aristarchus"
                Number of      Cross-         Cross-
Search Term     Citations     Citations      Citations
_____________   _________   ____________   _____________

"volcano"          196           54              4
"gas"             5881           24              1
"atmosphere"      4797           47              0
"eruption"         120            7              1
"flash"            262            1              0
"cloud"           1850           20              0
"nebulosity"        80            1              0
"mist"               5            0              0
"geyser"             5            0              0
"vapo(u)r"        1154            8              0
"transient"        458           15              3
"change"          4836           61              0
No First Term     ----         2440             10
________________________________________________________
\end{verbatim}

In contrast, during 1956-1968 there are 10 references to Aristarchus, seven
involving TLPs, all of which lead back to the Kozyrev reports.
Of these seven, five involve some of the TLP-related search terms that we have
used.
During this period TLPs are firmly fixed in many people's minds when discussing
particular lunar features, such as Aristarchus.

These papers (or lack thereof) can be considered an ``integral constraint'' on
the importance of observer preconception as to the existence of TLPs as an
important factor (for Aristarchus, at least) in determining the observation
selection function;
furthermore, before 1956 they provide no evidence for a ``hysteria signal'' of
false reports due to special attention.
Before 1956 TLP reports can be considered single, largely uncorrelated events,
and this partially justifies treating them with Poisson statistics.

This also implies that there must be some other reason that reports occur more
frequently at Aristarchus, by several orders of magnitude above what the area
ratio of $10^4$ km$^2$ to the near side surface of $6 \times 10^6$ km$^2$ would
imply.
The Elgers statement above implies that the ratio of observing time per area
for Aristarchus and the Plateau versus other areas not near the limb is at
least of order unity, and probably more.
However, in Crotts (2007a) we will see on the basis of alpha particle
transients from outgassing seen by Apollo and {\it Lunar Prospector} that TLPs
are correlated with $^{222}$Rn production and that this cannot with any
reasonable probability imply that TLPs occur over the entire Moon at the rate
reported near Aristarchus (and hence we are not simply being fooled because
human observers spend more time looking at the Aristarchus plateau).
Aristarchus is intrinsically more prone to TLPs, for some reason other than how
often people observe it.

\subsection{Statistically Consistent TLP Spatial Distribution}

While TLPs at Aristarchus events seem to be uncorrelated at least for
Aristarchus and for the period 1956 and before, we do not have sufficient
statistics for lesser TLP sites to perform the same tests.
I will simply assume that all events are Poisson.

I have not addressed how many reports are erroneous, and cannot
without more information about the TLP mechanism.
What I can address, however, is whether TLP report rates for various features
behave in statistically consistent manner.
In other words, what fraction of all reports is most justified for a particular
lunar feature, and is this fraction robust against changing a particular
variable?
We will choose variables which should be entirely unrelated to the conditions
of the lunar surface, but might might be tell-tale of other influences, for
example: ``irrelevant'' observer characteristics e.g., where they live, or
when they live.
If the TLP rate for a lunar feature depends on the location of the observer,
this may say more about the observational process than the physics of the Moon.
If the TLP rate for a feature depends on the historical period of observation,
this may imply changing influences on the observer (or it may indicate simply
that certain sites are more active at some time than at other, on historical
timescales).
Nonetheless, it may be instructive to construct a ``robust'' map of where on
the Moon TLPs are reported to originate.

Such a robust fraction for a feature is constructed by picking an ostensibly
random parameter (as it relates to lunar surface behavior), and then grouping
TLP reports according to various ranges in this parameter's value.
The robust fraction is computed by an average (usually not the
mean) according to some robust estimator e.g., the median.

The simplest robust estimate to construct is perhaps the median over
historical periods.
One-third (137, exactly) of unculled reports occur during the period 1892-1955,
with a roughly equal number before (134) and then after (145).
The resulting count as a function of feature-labelled ``pixel,'' analogous to
Table 1 (except that on average the current values are three times smaller), is
given in Table 5.
Specifically, we bin the counts seen in Figure 1 into 300 km square ``pixels''
and take the median count for each pixel from the three epochs.
Since each pixel can be labeled with the name of the feature(s) identified by
the observers in the reports that filled that pixel, we can list the corrected
count for each feature or group of features.
Within each pixel, we re-evaluate particular features to see if TLPs from the
two samples truly correspond geographically.
If TLPs occur in the same named feature (and we include any positional
information available), or within a 50 km radius of each other, or within
1.5$\times$ the radius of the named crater, whichever is larger, we retain this
as a match.
The latter is a rejection consideration in less than 10\% of the cases.
This resulting count from this entire procedure is likely to be much more robust
against selection biases than the distribution in Table 1, or for that matter
similar plots shown by previous authors who did not impose an artifact rejection
algorithm.
I am assuming in effect that there are quantitatively different observing
strategy results during these two time periods, which are capable of producing
spurious peaks in the geographic distribution of reports, but do not completely
neglect any area of the nearside Moon, excepting geometric effects such as limb
foreshortening or lunar phase selection due to evening/morning viewing times,
which are independent of time when averaged over the libration period (between
one day and one sidereal month).
My appraisal of the literature is that this is probably a good assumption.

\noindent
Table 5: Number of TLPs Reported Per Feature, Median of Three Historical Periods
\begin{verbatim}
__________________________________________________________________________
Robust
Report   Feature(s)
Count
-------  -----------------------------
46       Aristarchus/Schroter's Valley
13       Plato
 3       Kepler

2 each   Alphonsus, Eudoxus, Grimaldi, Mare Crisium, Posidonius

1 each   Alpetragius, Peak S. of Alps, Bessel, Calippus, Cassini, Carlini,
         Copernicus, Daniell, Gassendi, Godin, Hercules, Kant, La Hire,
         Littrow, Manilius, Mare Humorum, Mare Nubium, Messier, E. of Picard,
         SW of Pico, Proclus, Promontorium Heraclides, Ptolemaus, Riccioli,
         South Pole, Tycho
__________________________________________________________________________
\end{verbatim}

To within $1 \sigma$ Poisson errors, the contributions from the Aristarchus
region and Plato remain the same, at about 47\% and 13\% of the total.
Taken as a group, the large, young impact craters (Copernicus, Kepler, Tycho)
compose 5\%, consistent with the raw counts.
What is highly significant (at about the $4 \sigma$ level apiece) is the
disappearance of features Alphonsus, Gassendi and Ross D.
All of these were very prominent in post-1955 reports, but disappear in the
fraction of the robust count by factor of an order of magnitude or more.
Alphonsus, in particular, was one of the features that attracted greatest
attention following the report of Alter (1957) and Kozyrev.
Many of the observers in these latter reports were obviously aware of previous
observations, and in many cases were specifically targetting the crater because
of this.

Plato is a distinct, flooded crater on the northwestern edge of Mare Imbrium,
near mountainous regions such as Montes Alps, and appears very dark in
comparison.
It can be striking in its long shadows stretching across its face when near the
terminator.
Some observer descriptions sound suspiciously like reports of this normal
activity, but most do not correspond to normal appearance (see Haas 2003).
In 1854-1889 there were four reports involving at least some experienced 
observers noting extremely bright point sources that appeared for 30 min up to
5 h (the longest duration report we consider here); it is unclear if these
reports might have influenced each other.
There are few reports involving red sources (three not during eclipse); there
are many reports of cloud-like appearance.

Mare Crisium varies significantly in strength between this and the raw
result, and as we will see, between the different robustness estimates.
Since it is actually two ``pixels'' in diameter, I am unsure that this should
even be included as a feature in this analysis.

To illustrate the independence of the results on choice of historical period,
we consider other time intervals.
For instance, if we exclude the post-1955 period and slice the remaining sample
into three intervals (dividing the sample at 1877 and 1930), the median count
per 300 km square pixel, labeled by its primary feature, is shown in Table 6.

The values in Table 6 should be multipled by 4/3 in order to scale to Table 5.
There is little statistically significant change between the resulting report
counts, despite the complete exclusion of the post-1955 data.
Even if post-1955 is included in a robust (but non-median) average, as we present
in Table 7 and explain below (and in Crotts 2007a), the results are qualitatively
similar.
Even if pre-1956 data is time-sliced not in historical epoch, but time of the
year (January-April, May-Auguest, September-December), which would smooth out
any long-term fluctuations, the results are similar.

\newpage
\noindent
Table 6: Median Number of TLPs Reported Per Feature, Over Historical Periods
Pre-1956
\begin{verbatim}
__________________________________________________________________________
Median
Report   Feature(s)
Count
-------  -----------------------------------------------------------------
29       Aristarchus/Schroter's Valley
12       Plato
 5       Mare Crisium
 4       Tycho
 2       Kepler

1 each   Alphonsus, Peak S. of Alps, Bessel, Calippus, Cassini, Copernicus,
         Eudoxus, Gassendi, Godin, Grimaldi, Kant, Lichtenberg, Taurus
         Mountains, Macrobius, Messier, Picard, Pico, Posidonius, Proclus,
         Promontorium Heraclides, Ptolemaus, Riccioli, South Pole, Theaetetus
__________________________________________________________________________
\end{verbatim}

The two-sample robust estimator (Table 7) is constructed by taking the minimum
of the two values in equal-total samples.
This is useful in rejecting discrepant positive-going signals in cases where
one has only two copies of what should be otherwise identical images or maps,
but no good noise model (as is definitely the case here).
The fact that it rejects the same features as the all-history median (Table 5)
or the pre-1956 median (Table 6) suggests that the signals for Alphonsus,
Gassendi and Ross D might be systematic noise spikes that ride on an otherwise
roughly consistent data set for post-1955.
The lack of strong disagreement between the seasonal cut (Table 8) and all
others (Tables 5-7) might indicate that there are no strong historical episodes
of TLP activity for any of the strong features, since taking a timeslice
uniformly across the three centuries or more in data (by the seasonal
selection) yields the same result as slicing by historical period.
The only exception is the tiny interval 1956-1968 in which observational biases
are demonstrably different based simply on the correlations in the citation
record alone.
This explanation for the change in behavior is much easier to accept than a
sudden, simultaneous increase in lunar activity at Alphonsus, Gassendi and
Ross D.

\newpage
\noindent
Table 7: Number of TLPs Reported Per Feature, Comparing Pre- and Post-1930
Samples
\begin{verbatim}
__________________________________________________________________________
Robust
Report   Feature(s)
Count
-------  -----------------------------
66       Aristarchus/Schroter's Valley
15       Plato

2 each   Grimaldi, Messier

1 each   Alphonsus, Bessel, Cassini, Copernicus, Gassendi, Kepler,
         Lichtenberg, Littrow, Mare Humorum, Mare Nubium, Mons Pico, Pallas,
         Picard, Ptolemaeus, Riccioli, South Pole, Theaetetus, Tycho
__________________________________________________________________________
\end{verbatim}

\bigskip
\bigskip

\noindent
Table 8: Median Number of TLPs Reported Per Feature, Over Seasons of the Year
\begin{verbatim}
__________________________________________________________________________
Median
Report   Feature(s)
Count
-------  -----------------------------
25       Aristarchus/Schroter's Valley
13       Plato
5        Mare Crisium

2 each   Copernicus, Eratosthenes, Kepler, Tycho

1 each   Atlas, Bessel, Cassini, Grimaldi, Hansteen, Helicon, Herschel,
         Humboldt, Hyginus, Kant, La Hire, Lichtenberg, Messier, Picard,
         Pickering, Pierce A, Pico, Posidonius, Proclus, Promontorium
         Heraclides, Ptolemaeus, Riccioli
__________________________________________________________________________
\end{verbatim}

Finally, I slice the post-1955 sample in a non-temporal parameter, the location
of the observer at the time of the report.
There are roughly equal number of reports from three groups: Great Brittain,
continental western Europe, and the rest of the world (smaller by about 30\%,
and consisting primarily of the Americas and some reports in Asia and Russia).
The median of these, scaled to the same sample size is given in Table 9.
It shows qualitatively similar structure to all of the other robustness tests.

\noindent
Table 9: Median Number of TLPs Reported Per Feature, Varying Observer Location
\begin{verbatim}
__________________________________________________________________________
Median
Report   Feature(s)
Count
-------  -----------------------------
37       Aristarchus/Schroter's Valley
15       Plato
6        Mare Crisium
4        Tycho
2        Eratosthenes

1 each   Alphonsus, Atlas, Bessel, Calippus, Cassini, Copernicus, Gassendi,
         Godin, Grimaldi, Hercules, Kant, Kepler, La Hire, Lichtenberg,
         Macrobius, SW of Pico, Posidonius, Proclus, Promontorium Heraclides,
         Ptolemaeus, Riccioli, South Pole, Theaetetus
__________________________________________________________________________
\end{verbatim}

On the whole, however, the consistent behavior of the main features in the
sample lends credence to the notion that this approach has some validity.
We are testing whether given features are robust either in human observing
behavior, or in the long-term variability of the actual physical processes
producing TLPs at given sites.
At least we have varied the former in several significant ways and find its
effects to be consistent for most features, and inconsistent primarily in those
features where history casts some suspicion.

Mare Crisium is the only signal to vary significantly in strength between the
different robustness estimates.
Since it is actually two ``pixels'' in diameter, I am unsure that this should
even be included as a feature in this analysis.

I think that the results are sufficiently robust, even for Mare Crisium, to
allow one to average the relative frequencies in Tables 5-9 for various
features, summarized in Table 10.
A total of 49 features are listed (in order of decreasing statistical
significance, assuming Poisson behavior) in the Table; many of these are
undoubtably spurious.
Somewhat arbitrarily, we will require that the Poisson probability for a
feature's relative frequency excluding the value zero be greater than 48/49,
hence the detection be greater than about 2.4 $\sigma$.
Features that are at least this significant are Aristarchus/Schroter's Valley,
Plato, Mare Crisium, Tycho, Kepler, Grimaldi, and marginally Copernicus, which
together compose 74.3\% of the robust signal.

Table 10: Relative Frequencies of Robust TLP Reports by Feature
\begin{verbatim}
__________________________________________________________________________
Relative
Frequency &      Feature(s)
(1-sigma error)
---------------  -----------------------------
46.7% (3.3%)     Aristarchus/Schroter's Valley
15.6% (1.9%)     Plato
 4.1% (1.0%)     Mare Crisium
 2.8% (0.8%)     Tycho
 2.1% (0.7%)     Kepler
 1.6% (0.6%)     Grimaldi
 1.4% (0.6%)     Copernicus
[6.2% (1.2%)     sum of Tycho, Kepler & Copernicus]

 1.1% (0.5%)     Alphonsus, Bessel, Cassini, Messier, Ptolemaus, or Riccioli

 0.9% (0.5%)     Eratosthenes, Gassendi, Kant, Lichtenberg, (E. of) Picard,
                 (SW of) Pico (B), Posidonius, Proclus, Promontorium
                 Heraclides, or South Pole

 0.7% (0.4%)     Calippus, Eudoxus, Godin, La Hire, or Theaetetus

 0.5% (0.3%)     Peak S. of Alps, Atlas, Hercules, Littrow, Macrobius,
                 Mare Humorum, or Mare Nubium

 0.2% (0.2%)     Alpetragius, Carlini, Daniell, Hansteen, Helicon, Herschel,
                 Humboldt, Hyginus, Manilius, Pallas, Pickering, Pierce A,
                 or Taurus Mountains
__________________________________________________________________________
\end{verbatim}

It seems fairly clear that the behavior of major features in the distribution is
nearly bimodal.
Comparing, for instance the median in Table 5 versus the available total counts,
which should be three times larger, on average, the major features in Table 10
maintain nearly the average fraction of counts (one-third):
Aristarchus/Schroter's Valley: 46/150, and Plato: 13/45.
For the major, young impacts this is not so clear, for
Tycho$+$Kepler$+$Copernicus: $1+3+1$/$6+7+6$, and for Mare Crisium there are 15
total counts, which in consistent in some cases of robust counts and in some
cases not, marginally (which reduce anywhere from 0 to 6).
The maximum fraction of Grimaldi's counts are maintained: 2/4.
In contrast, the fractions for
Alphonsus is 2/22, Ross D: 0/13, and Gassendi: 1/18.
Taken together, $\sim$72/284, or 76\% of the expectation if all counts were
consistent except for random fluctuations.
One might consider this as evidence that most TLP reports are real (or at least
consistent), at least for sites with enough statistics to check.

These results show a surprising amount of regularity in the behavior of the TLP
sample, consistent at least with the possibility that many reports are real.
The spatial stucture, at least, is fairly consistent.
Now the question of the TLP mechanism must be addressed.
There are hypotheses as to possible non-lunar mechanisms that have been
advanced, plus reasons for why we should suspect TLPs in general.

\section{The TLP Controversy}

Any scientist should be skeptical of any conclusion based solely upon the
existing optical data base of TLP reports.
Most of them are anecdotal, not independently verified, and involve no
permanently recorded signal that did not pass through the human visual cortex.
Many of the observers are not professional, and some are not even very
experienced.
Undoubtably some, many, perhaps most of these reports are spurious.
Are they all spurious?
Is there any truth in these catalogs?
It is perhaps insufficient that a few special cases seem well-documented.
When selected from a huge data set, from all of the observers looking at the
brightest and most spectacular nighttime astronomical source, over a time
interval of four centuries, seemingly convincing random fluctuations will
occur.
One might despair that even with the robustness sieve implemented above that
an attempt to extract real information from this data set is likely to fail.

Perhaps the most condemning treatment of the TLP observations is Sheehan \&
Dobbins (1999).
This is not a scientific investigation but an essay in a popular magazine, and
is effective as such.
They propose without quantitative evaluation several arguments that cast doubt
on the reality of TLP reports:
1) several well-known cases of reported TLPs are suspect: Alter (1957), Kozyrev
(1962), Greenacre (1963) \& Barr, plus associated reports by the same
observers;
2) atmospheric refractive dispersion will cause red areas to appear shifted
with respect to points brighter than portions of the image around them;
3) the Corralitos Observatory TLP survey reported no positive detections;
4) the reports by Bartlett are spurious; and
5) TLPs are caused by transient optical mixing of bright and dark areas due to
bad seeing in front of areas of high source contrast.
Nonetheless, experienced observers aware of these criticisms are still
reporting TLPs and vouch in writing that these points do not explain their
observations.
Having laid criticism upon the TLP reports themselves, we consider these and
other possible objections with an open and critical eye, in order to understand
what may be occurring and whether we should believe it involves phenomena
arising near the lunar surface.

As treated above, any reasonable test of the Bartlett sample against the larger
TLP data base is likely to conclude that they are not both both chosen from the
same parent distribution.
One needs to hypothesize a different mechanism to explain this sample than the
one for other TLPs, and this cannot possibly be due to processes local to the
Moon.
We will not consider here the first two cases in (1), since they are
statistically insignificant.

The effect in (2) is accurately calculated given the report time, and latitude,
longitude and elevation of the observer, to a high degree of accuracy.
Sheehan \& Dobbins applies the effects of dispersion to the particular case of
Greenacre \& Barr, but it is unconvincing since (a) the same candidate event
was reported by observers two time zones away, where the airmass was much
different, and (b) the scale and direction of the dispersive color separation
corresponds only loosely to the reported observations.
To elaborate on the second point, at the time of observation (1963 October 30,
01:50 UT at mid-TLP), the Moon was approaching full at 11.9 days age, so that
Aristarchus was illuminated with the Sun about $16^\circ$ above the
horizon producing some shadows, cast to the WNW in lunar coordinates.
From the observer's viewpoint the Moon was at $66^\circ$ zenith distance, with
a parallactic angle of $-55^\circ$ dispersing red light (7000\AA) relative to
the eye's sensitivity peak ($\sim$5000\AA) 1.4 arcsec to the ENE in lunar
coordinates.
This is roughly the direction that would be needed to disperse the illuminated
rim of Aristarchus into shadow for red wavelengths, in the location observed
(to the SW of the crater).
What is not apparent is that a $1^{\prime\prime}.4$ displacement can produce
the extended region described, or the flux enhancement noted as ``brilliant''.

Does this mechanism explain the production of red TLPs, statistically?
There are 26 cases describing red or pink color, not during an eclipse, and for
which there is sufficiently accurate data to calculate a useable airmass.
The zenith distances for these reports, taken at mid-TLP observation, range
from $27^\circ$ to $80^\circ$, with a median of $56^\circ$, producing a
dispersive displacement between 5000\AA \ and 7000\AA \ of $0^{\prime\prime}.3$
to $3^{\prime\prime}.0$, median of $0^{\prime\prime}.8$ (or slightly less since
we do not correct for site elevation).
Subarcsecond displacements in the red, as most of these are, seem unlikely to
produce an observable effect.
Since the Moon never reaches more than $28^\circ .6$ from the celestial
equator, and most of these observing sites are far to the north ($+34^\circ$
to $+60^\circ$ latitude, with a heavy concentration to larger values), one
would expect few observations at zenith distance under $25^\circ$.
There is no particular tendency for red events to favor high airmass.
This explanation also has an intrinsic timescale of about 1-3 hours, which is
much longer then the reported timescale of at least 70\% of TLPs.
It is difficult to imagine a consistent source of modulation 10-100 times
faster.
In their book Sheehan \& Dobbins (2001) hypothesize a very rare phenomenon
relating to winter atmospheric considerations (that I am not sure I have
experienced in thousands of hours of observing, both with CCDs and visually),
but there is very little seasonal dependence in the red/pink TLP reports
(Jan.: 3, Feb.: 7, Mar.: 5, Apr.: 4, May.: 4, Jun.: 5, Jul.: 6, Aug.: 3,
Sep.: 6, Oct.: 9, Nov.: 9, Dec.: 7 -- almost completely in the northern
hemisphere, or suffering from ``bolivian winter'' in the north-central Andes).
Also, Sheehan \& Dobbins (1991) argue against a symmetric production by
atmospheric dispersion of blue TLPs, which should be displaced even more (by
about 20\%); I do not find their argument convincing but will not pursue this
here.

This leaves objection \#5: effects of seeing in high contrast source
distributions, and \#3: the absence of positive detections in the Corralitos
Observatory survey, the largest, probably most objective TLP search.
Considering seeing, this depends on the perception by the human visual system
which is hard to quantify without extensive psychological/physiological tests,
which are beyond the scope of this paper.
From my own observational experience, I think that an experienced human
observer would not be fooled by such an effect, whereas a novice observer might
be (see also Haas 2003).
This is particularly worrisome since this effect might be particularly in play
at regions of high surface brightness contrast, like the mare/highland
interface, or indeed at the crater Aristarchus.
(One might even worry that some inexperienced observers, noticing Aristarchus
for the first time, might report a TLP; there appear to be a few reports
consistent with this.)~
This does not explain why TLPs do not tend to be seen associated with many
small, bright points, usually fresh impact craters, across the otherwise
flatter mare visual field.
An electronic survey, using CCD or CMOS imaging detectors, could easily make
these effects moot, using established analysis techniques to compensate for
seeing variations.
We discuss this in detail in Paper III.

The Corralitos Observatory TLP survey spent some 8000 hours (10.9 months at
full duty cycle) observing, and was capable of covering the whole Moon in
15 min (although it is unclear if it always did), hence should have produced
some $3\times 10^4$ whole-Moon epochs.
What intrinsic event rate for TLPs should we assume?
As stated above, we might infer a rate of about once per month.
Given the structure of the distribution of observed TLP timescales, about 30\%
of the reports in the catalog would be missed.
If the Corralitos setup was equally sensitive as the typical TLP observer at
large, the absence of TLPs detected by them in the untriggered survey might
correspond to a $\sim 3 \sigma$ negative fluctuation (about 0.02\% Poisson --
not Gaussian -- random probability) in the expected counts {\it if}
observations were made at 100\% efficiency (which is unrealistic).

How sensitive was the Corralitos survey?
This is an important question that would have quantitative implications if we
knew the flux distribution function for TLPs, which we do not.
Still, the Corralitos was probably at least as sensitive as the typical TLP
observer and probably more so, so they should access at least the same event
rate.
The claimed sensitivity of the survey method seems improbably good:
better than a 5\% change in intensity in a 100\AA~band (Dunlap \& Hynek 1973),
corresponding to 0.5\%-1\% in a typical broad band characteristic of a
photometric optical color such as those employed, which then is converted to a
monochromatic, blinking contrast difference which is monitored by the eye.
This seems to be several times more sensitive than the threshold for the eye
detecting a constant monochromatic contrast, but the intent was apparently to
improve this threshold by causing the spot to blink at rate of a few Hz.
This may work for short periods of time, but the response of the eye to such a
signal fatigues over time in a manner that is most significant at rates of
about 12 Hz (Kanai \& Kamitani 2003).
Hynek et al.\ did their work before this effect had been studied scientifically
and it is unclear how they might have adjusted their observing procedure to
correct for this or perform tests to gauge the importance of such effects.
The likely effects of fatigue would need to be evaluated by reconstructing the
original setup of the Corralitos display equipment and this is difficult to
pursue.
In principle the same evaluation should be made of Moon Blink.

Particularly concerning are the TLP reports promptly transmitted to Corralitos
Observatory during the TLP patrol to provide confirmation or lack thereof.
Cameron (1978) lists 25 events that were apparently negative (four where this
is stated explicitly in terms of the data), two of these originating with
Bartlett are not included in our analysis, and two (\#1119 and \#1150) where
apparently Cameron disagrees with Hynek et al.\ and concludes there was a
positive confirmation.
This fraction of non-confirmation might lead one to conclude that many TLP
reports are not objectively real, at least in the midst of intensive campaigns
like those underway when these reports were produced (April 1966 - June 1969).

Despite the dedicated and laudable progress made by Cameron, Middlehurst,
Moore, Darling and centuries of researchers and observers, the current state of
the dataset resists application of the scientific method to the problem of
transient lunar phenomena.
There are striking examples of several well-documented cases where TLPs are
confirmed and suggest connection to physical mechanisms, but the strongest
evidence is anecdotal and leaves insufficient permanent records to allow the
testing and elaboration of hypotheses.
Given the transient nature of TLPs and state of available technology
heretofore, this outcome was difficult to avoid.

The onus of the argument must burden those who would convince us that TLPs are
real.
When it comes to locating a spurious effect that might explain the bulk of TLP
reports as unrelated to the vicinity of the Moon, absence of evidence is not
evidence of absence.
Given the inability heretofore to test a reported TLP in a timely manner with
sufficiently complementary measurements, we must ask if any other physical
effects firmly tied to the lunar environment are correlated with TLPs.

\section{Discussion, Summary and Conclusions}

A investigation by Cameron (1967, 1972) and Middlehurst (1977a, b) into
correlations with several possible lunar parameters turn up primarily null
relations e.g., lunar anomalistic period (time between perigees), and lunar age
(phase), and find some correlation with perigee and crossing of the Earth's
magnetopause and bow shock, plus a strong correlation with local sunrise which
might be a selection effect based on observers' attraction to this area of
higher contrast.
Middlehurst (1977a, b) also claims a statistically significant positional
correlation between TLPs and shallow moonquakes (from Nakamura et al.\ 1974),
which separately have been tied to $^{40}$Ar release (Hodges 1977, Binder
1980).

With the results in Section 2, we seem to have developed a reliable means to
compute the spatial distribution of the TLPs, and we should use this as a tool
for understanding their nature.
The sites of consistent TLP activity are dominated by Aristarchus and vicinity,
to a lesser extent Plato, followed perhaps by Mare Crisium, then the recent,
large impacts (Tycho, Kepler, Copernicus) and finally Grimaldi.
The often-reported sites Alphonsus, Gassendi and Ross D might be spurious.
In total, the area that is affected by such activity appears to be a vanishingly
small fraction of the lunar surface (at most a few percent, at least on the near
side).

In Crotts (2007a), however, we analyze the spatial distribution of non-optical
transient events on or below the lunar surface.
The robust distribution of TLPs found above corresponds to a striking
coincidence: of the four episodes when outbursts of $^{222}$Rn gas were detected
by virtue of alpha-particle detection (by detectors on {\it Apollo 15} and
{\it Lunar Prospector}, all correspond to the small minority of the lunar surface
responsible for the robust TLP reports (Grimaldi, Kepler and Aristarchus - twice).
Furthermore, there is a significant correlation between TLP loci and the edges of
maria, which is similar in description to the significant correlation between
maria edges and moonquakes (also in Crotts 2007a).
Also, this correlation with mare edges is seen for the density of alpha particles
from $^{210}$Po decay, which is a tracer for lunar outgassing as a product of the
decay of $^{222}$Rn gas.

Considering that 1) the spatial distibution of TLPs is robust across the lunar
near side regardless of various parameters tied to observer characteristics, and
2) this spatial distribution is highly correlated with tracers of lunar
outgassing ($^{222}$Rn and, indirectly $^{210}$Po), which we show elsewhere.
These two results greatly strengthen the case for the reality of TLPs: they
behave in a repeatable fashion, and they are tied to outgassing from the lunar
surface.

In Crotts \& Hummels (2007) and Crotts (2007b) we pursue this connection by
showing how outgassing from the lunar surface with produce TLPs and other effects
due to volatiles which might be studied to confirm or refute this picture, and
we also detail an array of measurement techiques which can further illucidate the 
TLP mystery, and tell us more about activity of lunar volatiles.
A key part of this effort is a robotic lunar imaging monitor, which is
practically capable of creating a new TLP data base without the enormous biases
present in the powerful but flawed human observer record.

\newpage
\section{(Appendix I) - A TLP Report Spanning Cislunar Space}

\noindent
Transcript of communications between {\it Apollo 11} and Capsule Communicator,
1969 July 19.

\noindent
Eight-digit numerical code: days, hours, minutes, seconds after launch
(JD 2440419.0639, geocentric reference).
Communicators are:\\
{\bf CC:} Capsule Communicator (CAP COMM)  Bruce McCandless\\
{\bf CDR:} Commander  Neil A.\ Armstrong \\
{\bf CMP:} Command module pilot  Michael Collins \\
{\bf LMP:} Lunar module pilot  Edwin E.\ Aldrin, Jr. 
\bigskip

\noindent
{\bf 03 04 56 35 CC:}~ Apollo 11, this is Houston. Over.

\noindent
{\bf 03 04 56 41 CDR:}~ Go ahead, Houston.

\noindent
{\bf 03 04 56 42 CC:}~ Roger. We show you, in the flight plan, staying in
orbital rate until about 79 hours 10 minutes.
Do you have some particular attitude or reason for wanting to go inertial?
Over.

\noindent
{\bf 03 04 57 00 LMP:}~ No, that's fine. I just wanted to confirm that. Until
79 10, then we'll breeze around here in orbit.

\noindent
{\bf 03 04 57 07 CC:}~ Roger. And we've got an observation you can make if you
have some time up there. There's been some lunar transient events reported in
the vicinity of Aristarchus. Over.

\noindent
{\bf 03 04 57 28 LMP:}~ Roger. We just went into spacecraft darkness. Until
then, why, we couldn't see a thing down below us. But now, with earthshine, the
visibility is pretty fair. Looking back behind me, now, I can see the corona
from where the Sun has just set. And we'll get out the map and see what we can
find around Aristarchus

\noindent
{\bf 03 04 57 54 CDR:}~ We're coming upon Aristarchus right now - -

\noindent
{\bf 03 04 57 55 CC:}~ - - Okay. Aristarchus is at angle Echo 9 on your ATO
chart. It's about 394 miles north of track. However, at your present altitude,
which is about 167 nautical miles, it ought to be over - that is within view of
your horizon: 23 degrees north, 47 west. Take a look and see if you see
anything worth noting up there. Over.

\noindent
{\bf 03 04 58 34 CDR:}~ Both looking.

\noindent
{\bf 03 04 58 36 CC:}~ Roger. Out.

\noindent
{\bf 03 05 03 01 LMP:}~ Houston, 11. It might help us a little bit if you could
give us a time of crossing of 45 west.

\noindent
{\bf 03 05 03 09 CC:}~ Say again, please, 11.

\noindent
{\bf 03 05 03 23 LMP:}~ You might give us a time of crossing of 45 west, and
then we'll know when to start searching for Aristarchus.

{\it [Note: the reader might want to skip this italicized, 9-minute section,
during which time the spacecraft approaches Aristarchus.]

\noindent
03 05 03 32 CC:~ Roger. You'll be crossing 45 west at 77 04 10 or about 40
seconds from now. Over. Thirty seconds from now.

\noindent
03 05 03 45 LMP:~ Okay.

\noindent
03 05 04 50 CC:~ Apollo 11, when we lose the S-band, we'd like to get
OMNI Charlie from you. And update my last, that 77 04 was the time when
Aristarchus should become visible over your horizon. 77 12 is point of closest
approach south of it. Over.

\noindent
03 05 05 14 LMP:~ Okay. That sounds better because we just went by
Copernicus a little bit ago.

\noindent
03 05 05 18 CC:~ Roger. We show you at about 27 degrees longitude right
now.

\noindent
03 05 05 25 LMP:~ Righto.

\noindent
03 05 07 07 LMP:~ Houston, when a star sets up here, there's no doubt about it.
One instant it's there, and the next instant it's just completely gone.

\noindent
03 05 07 16 CC:~ Roger. We copy.

\noindent
03 05 09 21 CC:~ Apollo 11, this is Houston. We request you use OMNI Charlie at
this time. Over.

\noindent
03 05 09 29 LMP:~ Okay. Going to OMNI Charlie.

\noindent
03 05 09 32 CC:~ Roger. Out.

\noindent
03 05 11 57 LMP:~ Houston, Apollo 11.

\noindent
03 05 12 01 CC:~ Apollo 11, this is Houston. Go ahead.

\noindent
03 05 12 06 LMP:~ Roger. Seems to me since we know orbits so precisely, and
know where the stars are so precisely, and the time of setting of a star or a
planet to so very fine a degree, that this might be a pretty good means of
measuring the altitude of the horizon ...
}

\noindent
{\bf 03 05 12 32 CC:}~ Roger.

\noindent
{\bf 03 05 12 51 CMP:}~ Hey, Houston. I'm looking north up toward Aristarchus
now, and I can't really tell at that distance whether I am really looking at
Aristarchus, but there's an area that is considerably more illuminated than the
surrounding area. It just has - seems to have a slight amount of fluorescence
to it. A crater can be seen, and the area around the crater is quite bright.

\noindent
{\bf 03 05 13 30 CC:}~ Roger, 11. We copy.

\noindent
{\bf 03 05 14 23 LMP:}~ Houston, Apollo 11. Looking up at the same area now and
it does seem to be reflecting some of the earthshine. I'm not sure whether it
was worked out to be about zero phase to - Well, at least there is one wall of
the crater that seems to be more illuminated than the others, and that one - if
we are lining up with the Earth correctly, does seem to put it about at zero
phase.
That area is definitely lighter than anything else that I could see out this
window. I am not sure that I am really identifying any phosphorescence, but
that definitely is lighter than anything else in the neighborhood.

\noindent
{\bf 03 05 15 15 CC:}~ 11, this is Houston. Can you discern any difference in
color of the illumination, and is that an inner or an outer wall from the
crater? Over.

\noindent
{\bf 03 05 15 34 CMP:}~ Roger. That's an inner wall of the crater.

\noindent
{\bf 03 05 15 43 LMP:}~ No, there doesn't appear to be any color involved in it, Bruce.

\noindent
{\bf 03 05 15 47 CC:}~ Roger. You said inner wall. Would that be the inner edge
of the northern surface?

\noindent
{\bf 03 05 16 00 CMP:}~ I guess it would be the inner edge of the westnorthwest
part, the part that would be more nearly normal if you were looking at it from
the Earth.

\noindent
{\bf 03 05 16 20 CC:}~ 11, Houston. Have you used the monocular on this? Over.

\noindent
{\bf 03 05 16 28 LMP:}~ Stand by one.

\noindent
{\bf 03 05 17 59 LMP:}~ Roger. Like you to know this quest for science has
caused me to lose my E-memory program, it's in here somewhere, but I can't find
it.\footnote{The E-memory was the spacecrafts erasable memory which held
temporarily programs for control of the spacecraft e.g., for guidance.}

\noindent
{\bf 03 05 18 08 CC:}~ 11, this is Houston. We're - we're hearing only a
partial COMM.
Say again please.

\noindent
{\bf 03 05 18 20 CDR:}~ I think ...

\noindent
{\bf 03 05 18 41 CDR:}~ Houston, we will give it a try if we have the
opportunity on next - when we are not in the middle of lunch, and trying to
find the monocular.

\noindent
{\bf 03 05 18 51 CC:}~ Roger. Copied you that time. Expect in the next REV you
will probably be getting ready for LOI 2.

\noindent
{\bf 03 05 19 09 CC:}~ So, let's wind this up, and since we've got some other
things to talk to you about in a few minutes. Over.

\bigskip
\noindent
{\bf Note:}
at the time of the above observation, the Moon's phase was 5.2d past new,
with Aristarchus in darkness, 26$^\circ$ from the anti-solar point and
57$^\circ$ from the sub-Earth point on the Moon.
The spacecraft was about 245 km in elevation above the lunar mean equatorial
surface and some 750 km from the center of Aristarchus, which appeared inclined
only 5$^\circ$ from edge-on.
Only the north-northwestern part of the inner rim would be easily seen.
At the time of the observation the phase angle was about 63$^\circ$, whereas
enhanced backscattering would be significant only for angles of a few degrees.

At mission elapsed time 03 05 14 (accurate to the minute), Pruss and Witte in
Bochum, GDR reported independently a 5-7 s brightening in Aristarchus (Cameron
1978 - we assign the longer timescale perhaps indicative of the {\it Apollo 11}
report, otherwise the Pruss \& White timescale is one of the shortest in the
catalog).
It is unclear precisely which event previous to this the Capsule Communicator
was indicating to {\it Apollo 11}.
He probably refers to a pulsing glow in Aristarchus reported by Whelan from
New Zealand some 12.3 h earlier.
That night there were nine TLP reports in our sample, primarily from LION (see
\S 2.4), with seven involving Aristarchus over a 14 h interval, including
independent reports (including photographs) over 03 05 58 -- 03 06 58 of
Aristarchus being brighter than normal.
There was no apparent attempt on {\it Apollo 11} to observe Aristarchus on the
next revolution, 2.15 h later in its initally wider and as yet uncircularized
lunar orbit. 
(The ``LOI 2'' burn -- Lunar Orbit Insertion -- at mission time 03 08 11 36
accomplished this circularization.)
The astronauts were busy, because the next day two of them would become the
first humans to walk on the Moon!

The point of this extensive excerpt to is illustrate a few important issues at
play in the data set, particularly in the interval around the Apollo era.
This is a unique example not only because of the setting, but because of the
degree to which the information flow is documented.
Is it a TLP report if observers are told to look at a specific area with
special attention?
Are the observers trained to distinguish the exceptional crater Aristarchus as
a spatial anomaly rather than a temporal one in comparison to other craters?
Do perhaps observers sometimes dismiss real temporal anomalies because they
have a mental model for normal appearance e.g., variations due to seeing -- or
in this case, the direct, 180$^\circ$ backscatter -- that might be caused by
less well-known effects?
To what extent can simultaneous, independent reports differ in description and
still be considered a confirmation?
Is it significant that many earlier selenographers made careful, repeated
observations with written records, or do more incidental observers provide
useful reports as well?

\noindent
{\bf Note on ``Flashes'' Observed from Lunar Orbit:}
Three instances of very rapid, bright flashes apparently from the lunar surface
were observed on {\it Apollo 16} (by Mattingly) and {\it Apollo 17} (separately
by Schmidt and Cernan).
While we do not analyze these in our sample, they are worth some separate
mention.
They are documented in the mission transcripts, debriefings, preliminary
science reports and in Cameron (1978).

The two {\it Apollo 17} reports were tied to Grimaldi and Mare Orientale,
respectively.
The first locus, and even the second (while indistinct), are sites of some of
the few outgassing events detected by means other than TLPs (both during
{\it Apollo 15}).
Grimaldi is a reasonably persistent TLP site, while Mare Orientale is too close
to the limb to be relevant.
This is interesting because there seem to have been very few if any of these
flashes seen {\it not} coming from the direction of the lunar surface, so
perhaps the explanation of them being caused by cosmic ray interactions with
the retina or vitrious humour of the eye is somewhat problematic.

The {\it Apollo 16} event location is more uncertain because it not only
occurred on the dark side, hence was difficult to localize visually.
Being on the far side, it cannot be tied to TLPs.
Nonetheless, it was reported by Mattingly as coming from below the horizon and
therefore austensibly the lunar surface.
As best as I can reconstruct from the available description, Mattingly was
looking in the vicinity of crater Korolev, on the far side.
This is highly uncertain.

\newpage
\section{References}
\noindent
``A Suspected Partial Obscuration of the Floor of Alphonsus''\\
\noindent
Alter, D.\ 1957, PASP, 69, 158.

\noindent
``Possibility of observing fall of meteorites on the moon from a station on the
earth''\\
\noindent
Anoshkin, V.A., Petrov, G.G.\ \& Mench, K.L.\ 1978, Astron.\ Vest., 12, 216.

\noindent
``Making the photos of flashes on the Moon''\\
\noindent
Arkhipov, A.V.\ 1991, Zemlia i Vselennaya, 3, 76.

\noindent
``The Bands of Aristarchus''\\
Barker, R.\ 1942, Popular Astronomy, 50, 192.

\noindent
``The bands of Aristarchus''\\
Barcroft, R.\ 1940, Popular Astronomy, 48, 302.

\noindent
``Lunar Flash Doesn't Pan Out''\\
\noindent
Beatty, J.L.\ 2003, Sky \& Tel., 105, 6, 24 (reporting results by
J.\ Westfall, and D.\ de Cicco \& G.\ Seronik).

\noindent
``Radio Emission of the Moon Before and After the Lunar Prospector Impact''\\
\noindent
Berezhnoy, A., Gusev, G.S., Khavroshkin, B.O., Poperechenko, A.B., Shevchenko,
V.V.\ \& Tzyplakov, A.V.\ 2000, in {\it Exploration and Utilisation of the
Moon, Proc.\ 4th Intern't'l Conf.\ Exploration \& Utilisation of the Moon},
eds.\ by B.H.\ Foing \&  M.\ Perry, (ESA SP-462), p.179

%
%
\noindent
``Shallow moonquakes - Argon release mechanism''\\
\noindent
Binder, A.B.\ 1980, Geophys.\ Res.\ Let., 7, 1011.

\noindent
``Correspondence - Meteors Seen in the Day-time''\\
\noindent
Bird, F.\ 1870, Astronomical Register, 8, 238.

\noindent
``The Lunar Craters Aristarchus and Herodotus''\\
\noindent
Birt, W.R.\ 1870, Astronomical Register, 8, 271.

%
%
\noindent
``Lunar Transient Phenomena: What Do the $Clementine$ Images Reveal?''\\
\noindent
Buratti, B.J., McConnochie, T.H., Calkins, S.B., Hillier, J.K.\ \&
Herkenhoff, K.E.\ 2000, Icarus, 146, 98.

\noindent
Buratti, B.J.\ 1999, at 31st Ann.\ D.P.S.\ Meeting, Abano Terme, Italy
(October 10-15).

\noindent
``Identification of the lunar flash of 1953 with a fresh crater on the moon's
surface''\\
\noindent
Buratti, B.J. \& Johnson, L.L.\ 2003, Icarus, 161, 192

%
\noindent
``Lunar crater Giordano Bruno - A.D. 1178 impact observations consistent with
laser ranging results''\\
\noindent
Calame, O.\ \& Mulholland, J.D.\ 1978, Science, 199, 875.

\noindent
``Operation Moon-Blink''\\
\noindent
Cameron, W.S.\ 1966, AJ, 71, 379.

\noindent
``Observations of changes on the Moon'' \\
\noindent
Cameron, W.S.\ 1967, 
Proc.\  5th Ann.\ Meet.\ Working Group on Extrater.\ Res., p. 47.

\noindent
``Comparative analyses of observations of lunar transient phenomena''\\
\noindent
Cameron, W.S.\ 1972, Icarus, 16, 339.

\noindent
``Lunar Transient Phenomena Catalog''\\
\noindent
Cameron, W.S.\ 1978, World Sp.\ Sci.\ Data Cntr., 78-03 (NSSDC, NASA-TM-79399).

\noindent
``Lunar transient phenomena''\\
\noindent
Cameron, W.S.\ 1991, Sky \& Tel., 81, 265.

%
\noindent
``Operation Moon Blink and report of observations of lunar transient phenomena''\\
\noindent
Cameron, W.S.\ \& Gilheany, J.J.\ 1967, Icarus, 7, 29.

\noindent
``Optical Observations of a Lunar Meteor Event during Leonid Meteor Showers in
2001''\\
\noindent
Chandrasekhar, T., Kikani, P.K.\ \& Mathur, S.N.\ 2003,
Bull.\ Astron.\ Soc.\ India, 31, 325.

%
%
%
\noindent
``Rate and Distribution of Kilogram Lunar Impactors''\\
\noindent
Cooke, W.J., Suggs, R.M., Suggs, R.J., Swift, W.R.\ \& Hollon, N.P.\ 2006,
LPSC, 38, 1986.

\noindent
``Lunar Outgassing, Transient Phenomena and The Return to The Moon, I:
Existing Data''\\
\noindent
Crotts, A.P.S.\ 2007a, Icarus (submitted).

\noindent
``Lunar Outgassing, Transient Phenomena and The Return to The Moon, III:
''\\
\noindent
Crotts, A.P.S.\ 2007b, ApJ (submitted).

\noindent
``An Automated Lunar Transient Imaging Monitor on Cerro Tololo''
\noindent
Crotts, A.P.S., Hickson, P., Pfrommer, T.\ \& Hummels, C.\ 2007 (in
preparation).

\noindent
``Lunar Outgassing, Transient Phenomena and The Return to The Moon, II:
''\\
Crotts, A.P.S.\ \& Hummels, C.\ 2007, ApJ (submitted).

\noindent
``The Status of Lunar Meteor Research (and Applications to the Rest of the Solar
System)''\\
\noindent
Cudnik, B.M.\ 2007, LPSC, 38, 1115.

\noindent
``Ground-Based Observations Of Lunar Meteoritic Phenomena''\\
\noindent
Cudnik, B.M., Dunham, D.W., Palmer, D.M., Cook, A., Venable, R.\ \&
Gural, P.S.\ 2003, Earth, Moon \& Planets, 93, 145-161.

\noindent
``Lunar Transient Phenomena Research Program: Ground Based Observing Programs,
Clementine Spacecraft''\\
\noindent
Darling, D.\ 2006,\\
\noindent
http://www.ltpresearch.org/programs/pastprograms/clementine\_spacecraft.htm

\noindent
``Langrenus: Transient Illuminations on the Moon''\\
\noindent
Dollfus, A.\ 2000, Icarus, 146, 430.	

%
\noindent
``The Corralitos Lunar Transient Phenomena (LTP) Surveillance Program
(1966-1972)''\\
\noindent
Dunlap, J.R. \& Hynek, J.A.\ 1973, BAAS, 5, 37

\noindent
``Lunar Leonids: November 18th Lunar Impacts''\\
\noindent
Dunlop, D.\ 1999,
http://iota.jhuapl.edu/lunar\_leonid/index.html.081299

\noindent
``First results from observations of the Moon by means of a polarimeter''\\
\noindent
Dzhapiashvili, V.P.\ \& Ksanfomaliti, L.V.\ 1962, {\it The Moon,
IAU Symp.\ 14}, (Academic Press: London)

%
\noindent
Ehrenfreund, P.\ 2006, {\it personal communication}

\noindent
``Selenographic Notes, February, 1884: Aristarchus amd Herodotus''\\
\noindent
Elgers, T.G.\ 1884, Astron.\ Register, 22, 39.

%
\noindent
``The Moon in Modern Astronomy (English Translation from German)''\\
\noindent
Fauth, Ph.\ 1906, (Owen: London).

%
%
%
%
%
\noindent
``Ejecta from lunar impacts: Where is it on earth?''\\
\noindent
Gault, D.E.\ \& Schultz, P.H.\ 1991, Meteoritics, 26, 336.

%
%
%
%
%
\noindent
Goodacre, W.\ 1899, Mem.\ Brit.\ Astron.\ Soc., 7, 52.

\noindent
{\it The Moon} \\
\noindent
Goodacre, W.\ 1931, (Pardy \& Son: Bournemouth, Eng.).

\noindent
``The 1963 Aristarchus Events''\\
\noindent
Greenacre, J.C.\ 1965, NYASA, 123, 811.

\noindent
``Was the Lunar Crater, Giordano Bruno, Formed on June 18, 1178?''\\
\noindent
Hartung, J.B.\ 1976, {\it Abstracts, Symposium on Planetary Cratering
Mechanics.}, LPI Contrib.\ 259, (Lunar Sci.\ Inst., Houston), p.43.

\noindent
``Lunar Changes in the Crater Aristarchus''\\
\noindent
Haas, W.H.\ 1938, Popular Astronomy, 46, 135.

\noindent
``Those unnumbered reports of lunar changes - Were they all blunders?''\\
\noindent
Haas, W.H.\ 2003, J.\ ALPO, 45, 2, 25.

%
\noindent
``Release of radiogenic gases from the moon''\\
\noindent
Hodges, R.R., Jr.\ 1977, Phys.\ Earth Planet.\ Interiors, 14, 282.

%
%
%
%
%
%
%
%
``Glazialkosmogenie''\\
\noindent
H\"orbiger, H.\ \& Fauth, P.\ 1913, (Unver\"anderter Neudruck, Leipzig).

%
\noindent
``A possible meteor shower on the Moon''\\
\noindent
Hunton, D.M., Kozlowski, R.W.H.\ \& Sprague, A.L.\ 1991,
Geophys.\ Res.\ Lett., 18, 2101.

\noindent
``A Search for Lunar Transient Phenomena: 1966-1968''\\
\noindent
Hynek, J.A.\ \& Dunlap, J.R.\ 1968, AJ, 73, 185

\noindent
``The Corralitos Observatory program for the detection of lunar transient
phenomena''\\
\noindent
Hynek, J.A., Dunlap, J.R.\ \& Hendry, E.M.\ 1976, NASA pub.\ CR-147888.

%
\noindent
``An Account of Two Voyages to New England''\\
\noindent
Josselyn, J.\ 1675, (London), p. 53 (2nd ed., 1865, p.45).

\noindent
``Time-locked Perceptual Fading Induced by Visual Transients''\\
\noindent
Kanai, R.\ \& Kamitani, Y.\ 2003, J.\ Cogn.\ Neurosci.\ 15, 664.

\noindent
``Photographic evidence of a short duration - Strong flash from the surface of
the moon''\\
\noindent
Kolovos, G., Seiradakis, J. H., Varvoglis, H.\ \& Avgoloupis, S.\ 1988, Icarus,
76, 525.

\noindent
``The origin of the Moon flash of May 23, 1985''\\
\noindent
Kolovos, G., Seiradakis, J.H., Varvoglis, H.\ \& Avgoloupis, S.\ 1992,
Icarus, 97, 142.

\noindent
``Observation of a Volcanic Process on the Moon''\\
\noindent
Kozyrev, N.A.\ 1959, Sky \& Tel., 18, 184.

\noindent
``Spectroscopic Proof for Existence of Volcanic Processes on the Moon''\\
\noindent
Kozyrev, N.A.\ 1962, in {\it Phys.\ \& Astron.\ of the Moon}, ed.\ Z.\ Kopal,
p.361.

%
%
\noindent
``Emergence of low relief terrain from shadow: an explanation for some TLP''\\
\noindent
Lena, R.\ \& Cook, A.\ \& 2004, J.\ Brit.\ Ast.\ Assoc., 114, 136.

%
%
\noindent
``Polarisation peculiarities of the crater Aristarchus''\\
\noindent
Lipsky, Yu.N.\ \& Pospergelis, M.M.\ 1966, Astronomicheskii Tsirkular, 389, 1.

%
\noindent
``Space debris and a flash on the Moon''\\
\noindent
Maley, P.D.\ 1991, Icarus 90, 326.

%
\noindent
``Are The Planets Inhabited?''\\
\noindent
Maunder, E.W.\ 1913, (Harper: London).

\noindent
``The Chinese Reader's Manual''\\
\noindent
Mayers, W.F.\ 1874 (Shanghai), p.\ 219

%
\noindent
``Lunar transient phenomena: Topographical distribution''\\
\noindent
Middlehurst, B.M.\ \& Moore, P.\ 1968, Science, 155, 449.

\noindent
``Transient Lunar Phenomena Observing Program: Lunar International Observers
Network''\\
\noindent
Middlehurst, B.\ 1970, in ``The Smithsonian Institution Center for Short-lived
Phenomena, Annual Report 1969'' (Smithsonian: Wash., DC), (transmitted by
D.\ Darling).

\noindent
``A survey of lunar transient phenomena''\\
\noindent
Middlehurst, B.\ 1977a, Phys.\ Earth Planet.\ Inter., 14, 185.

\noindent
``Transient lunar phenomena, deep moonquakes, and high-frequency teleseismic
events: possible connections''\\
\noindent
Middlehurst, B.\ 1977b, Phil.\ Trans.\ R.\ Soc.\ A.\, 285, 485.

\noindent
``Chronological Catalog of Reported Lunar Events''\\
\noindent
Middlehurst, B.M.,  Burley,, J.M., Moore, P.\ \& Welther, B.L.\ 1968, NASA
Tech.\ Rep.\ TR R-277.

\noindent
``The impact hazard''\\
\noindent
Morrison, D., Chapman, C.R.\ \& Slovic, P.\ 1994, in {\it Hazards Due to Comets
and Asteroids}, ed.\ T.\ Gehrels, (U.\ Arizona Press, Tucson).

\noindent
``Farside deep moonquakes and deep interior of the Moon''\\
\noindent
Nakamura, Y.\ 2005, J.\ Geophys.\ Res., 110, E01001. 

%
%
%
%
\noindent
``Medieval Chronicles and the Rotation of the Earth''\\
\noindent
Newton, R.R.\ 1972, (Johns Hopkins U.\ Press: Baltimore).

%
\noindent
``Detection of sporadic impact flashes on the Moon: Implications for the
luminous efficiency of hypervelocity impacts and derived terrestrial impact
rates''\\
\noindent
Ortiz, J.L., et al.\ 2007, Icarus, 184, 319.

\noindent
``Optical detection of meteoroidal impacts on the Moon''\\
\noindent
Ortiz, J.L., Sada, P.V., Bellot Rubio, L.R., Aceituno, F.J., Aceituno, J.,
Guti\'arrez, P.J.\ \& Thiele, U.\ 2000, Nature, 405, 921.

\noindent
``Are there at present Active Volcanos upon the Moon?''\\
\noindent
Pickering, W.H.\ 1892, The Observatory, 15, 250.

\noindent
``Visual Observations of The Moon and Planets''
\noindent
Pickering, W.H.\ 1900, Ann.\ Obs.\ Harvard Col., 32, 116.

\noindent
``The Moon: A Summary of the Existing Knowledge of Our Satellite, With a
Complete Photographic Atlas''
\noindent
Pickering, W.H.\ 1904a, (Doubleday: New York).

\noindent
``An Explanation of the Martian and Lunar Canals''
\noindent
Pickering, W.H.\ 1904b, Popular Astron., 12, 439.

\noindent
``Meteorology of the Moon''
\noindent
Pickering, W.H.\ 1916, Monthly Weather Rev., 44, 70.

\noindent
``The Moon flash of 1985 May 23 and orbital debris''\\
\noindent
Rast, R.H.\ 1991, Icarus 90, 328.

%
%
%
%
%
%
\noindent
``The TLP Myth: A Brief for the Prosecution''\\
\noindent
Sheehan, W.\ \& Dobbins, T.\ 1999, Sky \& Tel., 98, 3, 118.

%
%
%
\noindent
``A photo-visual observation of an impact of a large meteorite on the Moon''\\
\noindent
Stuart, L.H.\ 1956, J.\ ALPO, 10, 42.

\noindent
``The Historical Works of Gervase of Canterbury, Vol.\ 1''\\
\noindent
ed.\ Stubbs, W.\ 1879, (H.M.\ Stationery Off., London; repr.\ Kraus Repr.,
Ltd., 1965).

%
\noindent
``An Explosion on the Moon''\\
\noindent
Suggs, R.\ \& Swift, W.\ 2005,
http://science.nasa.gov/headlines/y2005/22dec\_lunar taurid.htm

\noindent
``A Meteoroid Hits the Moon''\\
\noindent
Suggs, R., Swift, W.\ \& Cooke, W.\ 2006,
http://science.nasa.gov/headlines/y2006/ 13jun\_lunarsporadic.htm

\noindent
``The Hiten Lunar Impact''\\
\noindent
Svedhem, H.\ 2006, in ``1st Internat'l Conf.\ on Impact Cratering in the Solar
System'' (ESTEC: Noordwijk).

%
%
%
\noindent
``The lunar radio flux during Leonid meteor showers and lunar eclipse''\\
\noindent
Volvach, A.E., Berezhnoy, A.A., Khavroshkin, O.B., Kovalenko, A.V.\ \&
Smirnov, G.T.\ 2005, Kin.\ Fiz.\ Neb.\ Tel. 21, 60.

%
%
\noindent
Veillet, C.\ 2006,
http://www.cfht.hawaii.edu/News/Smart1/ and
http://sci.esa.int/science-e/www/object/index.cfm?fobjectid=39961

%
\noindent
``The Lunar Craters Aristarchus and Herodotus''\\
\noindent
Whitley, H.M.\ 1870, Astronomical Register, 93, 194.

%
\noindent
``Meteor storm evidence against the recent formation of lunar crater Giordano
Bruno''\\
\noindent
Withers, P.\ 2001, Meteor.\ \& Planet.\ Sci., 36, 525.

\noindent
``The Moon in Ultra-Violet Light, and Spectro-Selenography''\\
\noindent
Wood, R.W.\ 1910, MNRAS, 70, 226

\noindent
``The first confirmed Perseid lunar impact flash''\\
\noindent
Yanagisawa, M.\ et al.\ 2006, Icarus, 182, 489.


\newpage

\begin{figure}
\plotone{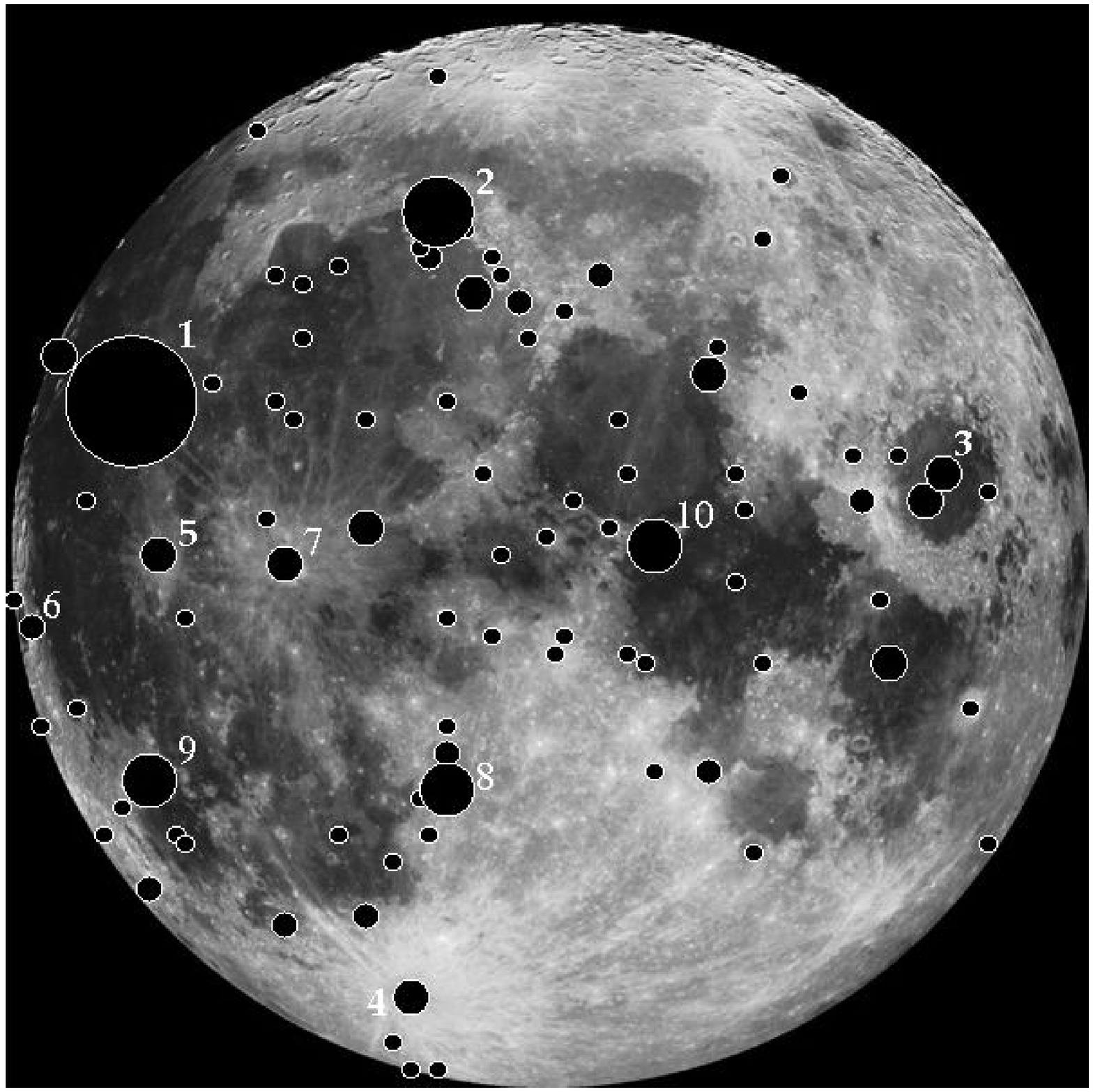}
\vskip 0.00in
\caption {
Distribution of TLP report loci as catalogued in Middlehurst et al.\ (1968),
with the exception of a minority of cases that are rejected for the reasons
detailed in the text.
The size of the symbols encodes the number of reports per features, as listed
in Table 1.
Marked features include: 1) Aristarchus (including Schr\"oter's Valley, Cobra's
Head and Herotus),
2) Plato,
3) Mare Crisium,
4) Tycho,
5) Kepler,
6) Grimaldi,
7) Copernicus,
8) Alphonsus,
9) Gassendi, and
10) Ross D.
The first seven features on the list survive with their samples nearly intact the
various robustness tests imposed, while the last three disappear nearly
completely. (Photo credit: NASA)
}
\end{figure}

\end{document}